\documentclass[twocolumn]{aastex62}
\usepackage{xcolor}

\submitjournal{ApJ}

\shorttitle{M dwarf Flares}
\shortauthors{Rodriguez Martinez et al.}

\newcommand{\ltsima}{$\; \buildrel < \over \sim \;$}
\newcommand{\simlt}{\lower.5ex\hbox{\ltsima}}

\def\arcdeg{\hbox{$^\circ$}}

\begin{document}

\title{A Catalog of M-dwarf Flares with ASAS-SN}

\correspondingauthor{Romy Rodr\'iguez Mart\'inez}
\email{rodriguezmartinez.2@osu.edu}

\author[0000-0003-1445-9923]{Romy Rodr\'iguez Mart\'inez}
\affil{Department of Astronomy, The Ohio State University, 140 W. 18th Ave., Columbus, Ohio 43210, USA}

\author{Laura A. Lopez}
\affil{Department of Astronomy, The Ohio State University, 140 W. 18th Ave., Columbus, Ohio 43210, USA}
\affil{Center for Cosmology and AstroParticle Physics, The Ohio State University, 191 W. Woodruff Ave., Columbus, OH 43210, USA}
\affil{Niels Bohr Institute, University of Copenhagen, Blegdamsvej 17, 2100 Copenhagen, Denmark}

\author{Benjamin J. Shappee}
\affiliation{Institute for Astronomy, University of Hawai'i, 2680 Woodlawn Drive, Honolulu, HI 96822, USA}

\author[0000-0002-7224-7702]{Sarah J. Schmidt}
\affiliation{Leibniz-Institute for Astrophysics Potsdam (AIP), An der Sternwarte 16, 14482, Potsdam, Germany}

\author[0000-0002-6244-477X]{Tharindu Jayasinghe}
\affil{Department of Astronomy, The Ohio State University, 140 W. 18th Ave., Columbus, Ohio 43210, USA}

\author{Christopher S. Kochanek}
\affiliation{Department of Astronomy, The Ohio State University, 140 W. 18th Ave., Columbus, Ohio 43210, USA}

\author{Katie Auchettl}
\affiliation{DARK, Niels Bohr Institute, University of Copenhagen, Lyngbyvej 2, 2100 Copenhagen, Denmark}
\affiliation{Department of Astronomy and Astrophysics, University of California, Santa Cruz, CA 95064, USA}

\author[0000-0001-9206-3460]{Thomas~W.-S.~Holoien}
\altaffiliation{Carnegie Fellow}
\affiliation{The Observatories of the Carnegie Institution for Science, 813 Santa Barbara St., Pasadena, CA 91101, USA}

\begin{abstract}

We analyzed the light curves of 1376 early-to-late, nearby M dwarfs to search for white-light flares using photometry from the All-Sky Automated Survey for Supernovae (ASAS-SN). We identified 480 M dwarfs with at least one potential flare employing a simple statistical algorithm that searches for sudden increases in $V$-band flux. After more detailed evaluation, we identified 62 individual flares on 62 stars. The event amplitudes range from $0.12 <\Delta V < 2.04$ mag. Using classical-flare models, we place lower limits on the flare energies and obtain $V$-band energies spanning $2.0\times10^{30} \lesssim E_{V} \lesssim 6.9\times10^{35}$ erg. The fraction of flaring stars increases with spectral type, and most flaring stars show moderate to strong H$\alpha$ emission. Additionally, we find that 14 of the 62 flaring stars are rotational variables, and they have shorter rotation periods and stronger H$\alpha$ emission than non-flaring rotational variable M dwarfs.  

\end{abstract}


\section{Introduction} 
\label{sec:intro}

Stellar flares are the consequence of surface magnetic fields. When magnetic field lines reconnect, they cause large, rapid flux increases in the UV, X-ray and sometimes optical wavelengths \citep{france13,jones:2016,hawley14}. Flares, sunspots and prominences, as well as other magnetic phenomena, have been extensively studied since the mid-1800s after the Carrington Event on the Sun \citep{carrington1859}. Subsequent studies of starspots, flare activity have been extended down to the low-mass end of the main sequence (e.g., \citealp{hawley96,kowalski13,hawley14,newton16,mondrik18,yang:2017,gunther19}).
Photometric observations of M dwarfs, cool and small stars with temperatures and masses between 2400--4000 K and 0.2--0.63 $M_{\odot}$, respectively \citep{gershberg05}, reveal strong chromospheric activity with starspots and flares and activity lifetimes that persist over Gyr timescales, longer than on Sun-like stars \citep{west2008}. 

Some open issues in stellar physics and activity include the evolution of the magnetic field strength and whether low-mass stars exhibit activity cycles like those observed in the Sun and Sun-like stars \citep{vida13, vida14}. Additionally, the dependence of flare rates on spectral type and age has not been fully characterized \citep{ilin18}. Previous studies suggest that M4 and later-type stars flare with higher frequency and larger amplitudes than earlier M stars \citep{kowalski09, davenport12, hawley14}. However, very low-mass stars ($< 0.35  M_{\odot}$) are typically convective and lack tachoclines \citep{chabrier97}, which raises questions about the drivers of heightened activity in the very low-mass regime. 

Other questions in stellar and flare physics involve the relationships between activity, ages and rotation rates. There is evidence that rapid rotators are more active than slow rotators (\citealp{kiraga07, newton17, mondrik18}),  but identifying and studying flares and activity cycles requires long-term observations of these stars. With the advent of large surveys like the All-Sky Automated Survey (ASAS; \citealp{pojmanski97}), Evryscope (e.g., \citealp{howard:2019}), the Panoramic Survey Telescope and Rapid Response System 
(Pan-STARRS; \citealp{kaiser04}), the \textit{Kepler} Space Telescope \citep{borucki10}, the Transiting Exoplanet Survey Satellite (\textit{TESS}; \citealp{ricker15}), and the All-Sky Automated Survey for Supernovae (ASAS-SN; \citealp{shappee14,kochanek17}), it has been possible to obtain detailed observations of flares across a wide range of spectral types and wavelengths, allowing some of these questions to be answered. However, our understanding of the rates and energy distributions of flares on M dwarfs is not complete, and an in-depth exploration of M-dwarf activity requires more observations. This is especially true in the very low-mass regime, where observations are more challenging because the stars are optically faint.

Here we search for flares from M-dwarf stars in ASAS-SN. ASAS-SN is an all-sky, optical survey with the primary goal of identifying bright supernovae and other transients. ASAS-SN presently consists of twenty 14-cm telephoto lenses, covering around 16000 sq. degrees at a median cadence of roughly 21 hours in g band and $\sim$2$-$3 days in $V$ band from 2012 to 2018 (the time span considered in this paper).   

Several strong flares have been found with ASAS-SN \citep{stanek13,simonian16,rodriguez18} and characterized in detail \citep{schmidt14,schmidt16}. Recently, \citet{schmidt18} compiled a catalog of M-dwarf flares serendipitously identified in the first four years of ASAS-SN transient alerts. \citet{schmidt18} followed-up these events with additional photometry and spectroscopy of the host stars. 

The study of flares and stellar activity of M dwarfs is important for the habitability of planets orbiting these small, cool stars. Transit and radial velocity exoplanet missions have uncovered an abundance of small planets in the habitable zones of M dwarfs (e.g., \citealp{dressing15}), making them crucial targets in the search for habitable worlds. However, planets in the habitable zones of M dwarfs are more exposed to stellar activity, including strong X-ray and UV emission from flares, because the habitable zones are only $\sim$0.1 AU from the star. As such, the activity of these stars is important because planetary habitability depends on both intrinsic planet properties and the characteristics of their host stars. 

Although previous studies suggest that stellar activity negatively affects planetary atmospheres and potential surface life, the details remain an open question (e.g., \citealp{segura10,davenport16,vida17,omalley18}).  Recent observations of strong and frequent flares in the mid-M dwarf Proxima Centauri have raised doubts about the existence of an atmosphere and therefore the habitability of its Earth-mass exoplanet \citep{davenport16, howard18}. There is significant evidence suggesting that magnetic activity and flares can cause atmospheric erosion \citep{lammer07}, runaway greenhouse effects, and hydrodynamic escape of the atmospheres \citep{luger15,shields16}.

Large-scale studies of activity and flares on nearby, planet-bearing M dwarfs of all spectral types will allow prioritization of systems for follow-up observations with James Webb Space Telescope (\textit{JWST}) and other upcoming missions. Stars with low-activity are more promising targets for {\it JWST} if the aim is to find habitable exoplanets. Detailed observations of active planet hosts can reveal how activity affects planet atmospheres, and probe other relevant problems, like the interaction between stellar and planetary magnetic fields (e.g., \citealt{poppenhaeger15}).

Flares are unpredictable transients, so a systematic study of flares requires frequent observations over extended periods of time. In this paper, we examine two samples of M dwarfs using ASAS-SN.  We search for flares in the magnitude-limited sample of earlier M-dwarfs from \citet{lepine13} and in the volume-limited ($<$ 20 pc) sample of later M-dwarfs from \citet{cruz07}.  After identifying flares, we investigate the correlations between flare frequency, flare energy, spectral type and H$\alpha$ as a stellar activity indicator. 

The structure of this paper is as follows. In Section~\ref{sec:data}, we describe the sample and our methodology to identify flares. We also discuss the potential sources of false positives and completeness. In Section~\ref{sec:energies}, we estimate the energies of the identified flares, and in Section~\ref{sec:rotation}, we identify rotational variables in our sample and measure their rotational periods. In Section~\ref{sec:habitability}, we note which candidates in our sample host confirmed or potential exoplanets. Finally, in Section~\ref{sec:conclusions}, we summarize our results.

\begin{deluxetable}{l c c}[bt]
\tabletypesize{\small}
\tablecaption{Spectral types for the stars in our sample \label{table:spectraltypes}}
\tablewidth{0pt}
\tablehead{
\colhead{spt} & \colhead{\citet{lepine13}} & \colhead{\citet{cruz07}}}
\startdata
M0 & 337 & -\\
M1 & 299 & 1\\
M2 & 260 & 1\\
M3 & 250 & 2\\
M4 & 108 & 1\\
M5 & 18 & 11\\
M6 & 2 & 20\\
M7 & 1 & 31\\
M8 & 1 & 18\\
M9 & - & 10 \\
\hline
\enddata
\end{deluxetable}

\section{Observations and Data Analysis} \label{sec:data}

\subsection{The Sample}
\label{sec:sample}

The majority of the sources we analyzed were taken from the 
\citet{lepine13} sample of bright M dwarfs. \citet{lepine13} is a spectroscopic, magnitude-limited catalog of the brightest ($J<9$ mag) late-K and M dwarfs in the Northern hemisphere. It is comprised of 1564 sources, of which 1408 are spectroscopically confirmed to be stars between K7 and M6 in spectral type. This catalog is estimated to contain $\sim$90\% of all M dwarfs in the North with $J<9$ mag and they have a mean $V$-band brightness of $V\sim12$ mag. The spectral classifications have an estimated precision of a half-subtype. The dwarf classifications require proper motions of $\mu > 40$ mas yr$^{-1}$ to exclude red giants. We used only the sources with reliable spectral classifications, leading to a total of 1276 objects between M0 and M8 (see Table~\ref{table:spectraltypes}).

\citet{lepine13} estimated the effective temperatures of these stars by fitting their spectra with PHOENIX atmospheric models \citep{allard10}. They also identified active M dwarfs based on their $\rm H\alpha$ features. The strength of this line is a standard diagnostic of chromospheric activity in M dwarfs, with the most active M dwarfs showing H$\alpha$ in emission rather than absorption, although some active late-K and early M dwarfs can also show H$\alpha$ in absorption \citep{walkowicz09,newton17}.  These M dwarfs have 2660 K $< T_{\rm eff} <$ 3940 K and a typical $\rm H\alpha$ equivalent width of $-3$ $\rm \AA$. 

We complement these sources with a volume-limited ($< 20\rm$ pc) sample of late-M and ultracool dwarfs identified from 2MASS by \citet{cruz07}. The \citet{lepine13} and \citet{cruz07} samples are different and do not overlap. As with the \citet{lepine13} sample, we only use the objects spectroscopically classified as M dwarfs, yielding a sample of 100 M dwarfs with spectral types between M1--M9 (see Table~\ref{table:spectraltypes}), of which 97\% are M5 or later. These sources have typical $V$-band magnitudes of $V\sim 16$ mag. The stars from both catalogs were selected without reference to H$\alpha$ emission or other classifiers of magnetic activity, so the sample includes both active and inactive stars.

Although we did not use this information while vetting the candidates, there is external data on the activity of these stars. In Simbad \citep{wenger:2000} and the International Variable Stars Index (VSX, \citealp{watson06}), 122 are identified as flare stars, generally UV Ceti stars which show flares with $\Delta V \simeq 0.1$ to  $\Delta V \simeq 6$ mag. The others with flare star designations in Simbad come from a sample of $\sim$570 stars flagged in the GALEX all-sky UV survey \citep{jones:2016} because they have excess UV emission. 

\subsection{The All-Sky Automated Survey for Supernovae (ASAS-SN)}
\label{sec:asassn}

After selecting the candidates, we obtained ASAS-SN aperture photometry for each of the 1376 stars, and we modified the standard ASAS-SN photometry pipeline to account for proper motions of the M dwarfs. Prior to 2018, ASAS-SN consisted of 4 telescopes monitoring the sky every 2$-$3 days to $V\simeq 17$. Afterwards, it rapidly built up to 20 telescopes monitoring the sky every 21 hours to $g\simeq18$. ASAS-SN takes 3 dithered 90-second exposures, and we obtained the light curves for these individual exposures rather than the sum. ASAS-SN has magnitude limits of $V\approx16.5-17.3$ and $g\approx18$ and photometric precision of $\sim$0.02 mag at $V \sim$12 mag and $\sim$ 0.08 mag at $V \sim$16 mag \citep{jayasinghe18}.

One disadvantage of ASAS-SN is that its cadence is low and the integration time of 90 seconds is short relative to the timescale of a flare ($\sim$100 seconds; \citealt{davenport16}). This limits both the number of detections of a given flare, as well as the number of observable flares from each star. The former means that the shape or morphology of the flare, with its characteristic fast rise and slower exponential decay, is not well constrained. This is a limitation compared to large-scale flare searches on M dwarfs with high-cadence photometry using, for example, \textit{Kepler}/K2 \citep{davenport16,doyle18} or Evryscope \citep{howard18,howard:2019}.

\subsection{Flare identification}
\label{sec:flare ids}

\begin{figure*}[t]

\begin{center}
\includegraphics[width=0.9\textwidth]{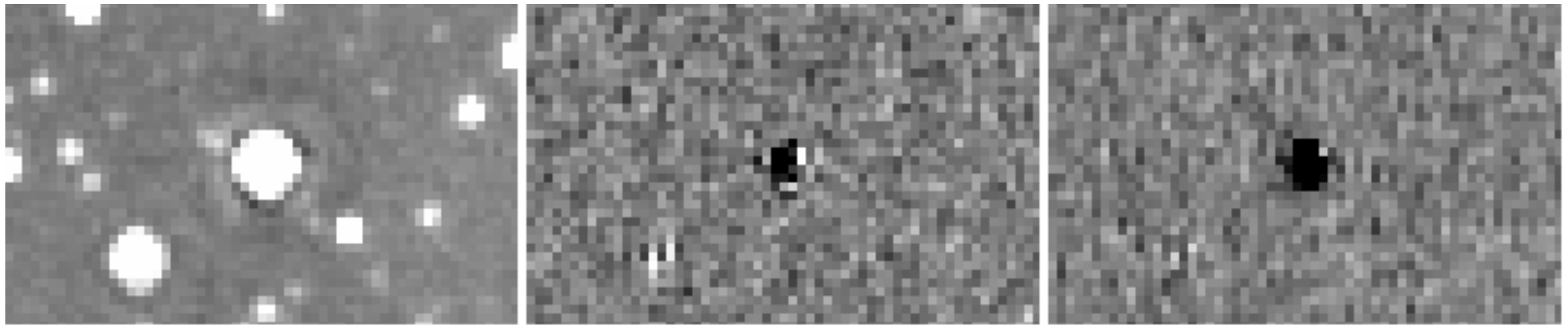}
\end{center}
\vspace{-3mm}
\caption{A $\Delta V = 0.14$ mag flare on the early-M star I04480+1703. The target lies at the center of each image. The image on the \textbf{left} is the ASAS-SN reference image of the star, while the \textbf{middle} and \textbf{right} images are the difference between the reference and two images at the epoch of flare.}
\label{fig:fig1}
\end{figure*} 

In order to identify flares, we searched all of the M-dwarf light curves to select the sources with potential flares. We computed the mean, $V_{\rm mean}$, median $V_{\rm median}$, and standard deviation $\sigma$ of each light curve. We then identified positive magnitude excursions that were $>2\sigma$ from the light curve and selected events that satisfied
\begin{equation}
    \frac{V_{\rm peak}-V_{\rm median}}{\sigma} \geq 2
\label{eq:1}
\end{equation}
where $V_{\rm peak}$ is the peak magnitude of the flare. We used the median rather than the mean because the former is less sensitive to outliers. We chose a statistical cut of $2\sigma$, in contrast to other works who use slightly higher cuts (e.g., \citealt{kowalski09} and \citealt{hawley14} use 3$\sigma$ and 2.5$\sigma$, respectively) to try to identify smaller flux increases.   

Many of the $\geq$ M5 stars in the \citet{cruz07} sample are not directly detected by ASAS-SN and have light curves that largely consist of upper limits. For these stars we define a median magnitude based on the number of upper limits and good detections. If a light curve consisted only of upper limits, we discarded it, regardless of whether it had statistically significant outliers, as upper limits are not reliable enough for the confirmation of flares. If the light curve had a combination of upper limits and detections, then we took the median of the entire light curve. Finally, if it consisted of mostly detections (i.e., not upper limits), then we only considered the detection median rather than the median on the entire light curve. 

After applying these criteria, we had 556 stars with at least one candidate flare in the \citet{lepine13} sample and 27 in the \citet{cruz07} sample. Many of these candidates were false positives from 2012 when ASAS-SN had just started, so we dropped this early data. We also excluded saturated stars (with $V < 10$ mag) since the corrections for saturation in the ASAS-SN pipeline (see \citealt{kochanek17}) are not reliable. After rejecting images taken in bad weather, with FWHM $\geq$ 2.5 pixels (71 candidates) and saturated stars (32 candidates), 453 stars with candidate flares were left from \citet{lepine13} and 27 from \citet{cruz07}.

Next, we inspected the light curves to confirm or discard the candidates. In particular, we examined the subtracted images of the candidate flares.  If the flare candidate is real, the subtracted image should show a negative, star-like image at the location of the source, such as that shown in Figure~\ref{fig:fig1}. In general, we kept flares that had at least 2 consecutive, $> 2\sigma$ detections of the flare and a clear signal in the subtraction images at the time of the flare. In $\sim$10 cases, we kept flares with only one $\geq$ 2$\sigma$ detection because the signal in the subtracted image was strong and clean.

Consecutive points that were marginally above the threshold (i.e., a few millimags above the threshold) were not considered as flares because they likely result from some systematic problem. We also verified that the flare durations of the putative events were consistent with the durations of well-studied classical (singly-peaked) flares. Almost all observed M-dwarf flares have durations of less than 90 minutes \citep{hawley14}, although there are exceptions (e.g., \citealp{kowalski10}). None of the flares identified in our sample had durations of more than 90 minutes.

We assume that all confirmed events are classical flares rather than a mixture of classical and complex (multi-peaked structures; see e.g., \citet{hawley14} for a discussion of classical and complex flares). Although this assumption is not necessarily true, ASAS-SN does not have the photometric precision and cadence to resolve multi-peaked events, especially since these can last longer than classical flares (up to $\sim$8-12 hours, e.g, \citealp{kowalski10}).

At the end of the validation process, we had a final sample of 62 stars with at least one flare in the \citet{lepine13} sample. We could not confirm any of the candidates in the \citet{cruz07} sample. Most of these stars are near the detection limit of ASAS-SN ($V\approx16.5-17.3$ mag), undetected in their quiescent state and with only a single detection during the putative flare. The flare images for these sources were also not of high quality compared to those from the \citet{lepine13} sample. Consequently, the data were only sensitive to rarer, higher amplitude flares from a small number of stars.

We designate 8 flares as ``Maybes", where the candidates could not be confirmed (see Table~\ref{tab:stellarprops}). A few of these are from the early ($< 2013$) phase of ASAS-SN, when systematic issues were less well controlled.

\begin{deluxetable*}{lccccccc}[t]
\tablenum{3} \tablecolumns{8}
\tablecaption{Stellar Properties\label{tab:stellarprops}}
\tablehead{
\colhead{ID} & \colhead{Spt} & \colhead{$\zeta$} & \colhead{log$g$} & \colhead{$T_{\rm eff}$ (K)} & \colhead{Parallax\tablenotemark{a} (mas)} & \colhead{Distance (pc)} &  \colhead{H$\alpha$ EW  (\AA)}}
\startdata
\cutinhead{Confirmed Flares}
I00118+2259 &  M3.5 & 1.10 & 4.5& 3260 & $48.860\pm0.080$ & $20.45^{+0.03}_{-0.03}$ & +0.30  \\
I00162+1951W & M4.0 & 1.06 & 5.0 & 3250 & $4.290\pm0.040$& $15.18^{+0.02}_{-0.02}$& $-$4.01 \\
I00325+0729 & M4.0 & 1.05 & 4.5 & 3150 & $28.089 \pm 0.155$& $35.56^{+0.19}_{-0.19}$ & $-$3.97 \\
\textbf{I01033+6221} & M5.0 & 1.00 & 4.5 & 2940 & $101.637 \pm 0.080$& $9.83^{+0.007}_{-0.007}$&$-$8.06 \\
\textbf{I01593+5831} &  M4.0 & 1.04 & 5.0 & 3290 & $76.13\pm  0.054$ & $13.12^{+0.009}_{-0.009}$ &$-$5.00 \\
\textbf{I02002+1303} &  M5.0 & 0.99 & 5.0 & 3130&  $223.634 \pm 0.106$ &  $4.47^{+0.002}_{-0.002}$ &$-$0.83\\
\textbf{I02088+4926} & M4.0 & 0.93 & 5.0 & 3270 & $58.598\pm0.078$ &  $17.05^{+0.02}_{-0.02}$ &$-$3.76 \\
\textbf{I02336+2455} &  M4.5 & 1.03 & 5.0 & 3130 & $0.579\pm 0.159$ &  $9.96^{+0.01}_{-0.01}$ &$-$3.06\\
\textbf{I03394+2458}  & M3.5 &  0.97 & 5.0& 3380 &$62.542\pm 0.079$&  $15.98^{+0.02}_{-0.02}$ &$-$0.65\\
I04284+1741 &  M2.0 & 0.95 & 5.0 & 3590 &$21.817\pm 0.231$ & $46.83^{+0.58}_{-0.56}$ &$-$1.95\\
I04480+1703 &  M0.5 & 0.92 & 4.5 & 3730 & $22.650\pm 0.451$ & $44.12^{+0.90}_{-0.86}$ & +0.07\\
I05019+0108$^\star$ &  M4.5 & 1.00 & 5.0 & 3160 & $39.554\pm 0.083$ &  $25.26^{+0.05}_{-0.05}$ &$-$3.89\\
I05062+0439$^\star$ & M4.0 & 1.01 & 4.5 & 3100 & $35.966\pm0.054$ & $27.78^{+0.04}_{-0.04}$ &$-$5.84\\
\textbf{I05091+1527} & M3.0 & 1.02 & 5.0 & 3520 &$33.617\pm0.0730$ & $29.72^{+0.06}_{-0.06}$ &$-$1.80\\
\textbf{I05337+0156} &  M3.0 & 0.85 & 4.5 &3350 &$63.598\pm0.088$ &$15.71^{+0.02}_{-0.02}$ &$-$4.98\\
I05547+1055$^\star$ &  M3.0 & 0.90 & 5.0 &3530 & $40.520\pm0.047$ &$24.66^{+0.02}_{-0.02}$ &$-$3.55\\
I07364+0704 &  M5.0 & 0.96 & 4.5 &2950 & $149\pm 44^{\dagger}$ & $6.71\pm 1.98$  &$-$3.55\\
I07384+2400$^\dagger$ &  M3.5 & 0.89 & 5.0 &3270 &$51.590 \pm 0.058$ & $19.37^{+0.02}_{-0.02}$ &$-$4.08\\
\textbf{I07558+8323} &  M4.0 & 0.98 & 5.0 &3250 & $76.352 \pm0.035$& $13.09^{+0.006}_{-0.006}$ &$-$3.27\\
\textbf{I08589+0828} &  M4.5 & 0.95 & 5.0 &3140 & $215\pm 63^{\dagger}$ & $4.65\pm 1.36$ &$-$2.74\\
I09177+4612 &  M2.5 & 0.92 & 4.5 &3400 &$29.898 \pm0.424$& $33.42^{+0.48}_{-0.47}$ &$-$2.89\\
\textbf{I10360+0507} &  M4.0 & 0.91 & 5.0 &3140 &$65.381 \pm0.091$& $15.28^{+0.02}_{-0.02}$ &$-$2.80\\
I10367+1521$^\star$ &  M4.5 & 0.92 & 5.0 &3100 &$50.621\pm 0.157$& $19.74^{+0.06}_{-0.06}$&$-$4.46\\
I11033+1337 &  M4.0 & 0.90 & 5.0 &3190 & $65.59\pm 0.09$ & $15.23^{+0.02}_{-0.02}$ &$-$1.06\\
\textbf{I12189+1107} &  M5.0 & 1.07 & 5.0 &3110 &$154.507\pm 0.110$ &$6.47^{+0.004}_{-0.004}$ &$-$3.53\\
I12332+0901 &  M5.5 & 1.00 & 4.5 &2850 & $258 \pm 76^{\dagger}$ & $3.87\pm 1.14$ & $-$4.51\\
I12485+4933$^\star$ &  M3.5 & 0.98 & 5.0 &3370 &$40.771\pm 0.248$ &$24.51^{+0.15}_{-0.14}$ &$-$4.73\\
\textbf{I13007+1222} &  M1.5 & 0.91 & 4.5 &3570 &$86.856\pm0.151$& $11.50^{+0.02}_{-0.02}$ &$-$1.96\\
I13317+2916 &  M4.5 & 0.98 & 5.0 &3150 &$54.687\pm0.331$ &$18.27^{+0.11}_{-0.11}$ &$-$7.51\\
I15126+4543 &  M4.0 & 1.01 & 5.0 &3270 &$64 \pm 19^{\dagger}$ & $15.62\pm 4.63$ &$-$3.19\\
\textbf{I15218+2058} &  M2.0 & 0.95 & 4.5 &3490 &$87.378 \pm0.049$ &$18.27^{+0.006}_{-0.006}$ &$-$2.25\\
I15238+5609 &  M1.0 & 0.91 & 4.5 &3570 &$19.657 \pm0.0260$ &$50.79^{+0.06}_{-0.06}$ &$-$1.97\\
\textbf{I15555+3512} &  M4.5 & 1.02 & 5.0 &3130 &$35.942 \pm 0.0440$ &$27.80^{+0.03}_{-0.03}$ &$-$5.72\\
I15557+6840$^{\star}$ &  M2.5 & 0.98 & 5.0 &3520 &$38.995 \pm 0.024$ &$25.62^{+0.01}_{-0.01}$ &$-$2.84\\
\textbf{I15581+4927}$^\star$ &  M1.0 & 0.86 & 4.5 & 3520 &$26.349 \pm 0.021$ & $37.91^{+0.03}_{-0.03}$ &$-$2.95\\
I16328+0950 &  M3.5 & 0.97 & 5.0 & 3270 & $65.047 \pm 0.065$& $15.36^{+0.01}_{-0.01}$ &+0.31\\
\textbf{I17198+2630} &  M3.5 & 1.02 & 4.5 &3260 & $92.966 \pm0.061$& $10.75^{+0.007}_{-0.007}$ &$-$0.35\\
I17338+1655$^\star$ &  M5.5 & 1.02 & 5.0& 2940 & $62.371\pm 0.390$& $16.02^{+0.10}_{-0.10}$ &$-$9.16\\
I18022+6415 &  M5.0 & 0.97 & 4.5& 3040 & $92\pm27^{\dagger}$  &$7.78^{+0.003}_{-0.003}$ &$-$3.17\\
\textbf{I18358+8005} &  M4.0 & 0.95 & 5.0& 3280& $60.959\pm0.041$ & $16.39^{+0.01}_{-0.01}$ &$-$2.32\\
\textbf{I18427+1354} &  M4.5 & 0.97 & 5.0 & 3100 & $91.429\pm 0.070$& $10.93^{+0.008}_{-0.008}$ &$-$2.26
\enddata
\tablenotetext{a}{Targets in bold are classified as Flare Stars or UV Ceti-type stars in Simbad or in VSX. Objects with a $^{\star}$ are rotational variables. Parallax values with a $^{\dagger}$ come from the photometric parallaxes reported in \citet{lepine13}, while those without the $^{\dagger}$ are from \textit{Gaia} DR2. The $\zeta$, $\zeta_{\rm TiO/CaH}$ or ``zeta" parameter is measured and defined in \citet{lepine13} as a combination of the TiO5, CaH2 and CaH3 spectral indices, a quantity shown to be correlated with metallicity in metal-poor, M subdwarfs and is $\zeta \simeq 1.05$ for solar abundances.}
\end{deluxetable*}

\begin{deluxetable*}{lccccccc}[ht]
\tablenum{3} \tablecolumns{8}
\tablecaption{Stellar Properties (continued)\label{tab:stellarprops}}
\tablehead{
\colhead{ID} & \colhead{Spt} & \colhead{$\zeta$} & \colhead{log$g$} & \colhead{$T_{\rm eff}$ (K)} & \colhead{Parallax\tablenotemark{a} (mas)} & \colhead{Distance (pc)} & \colhead{H$\alpha$ EW  (\AA)}}
\startdata
I19146+1919  & M3.5 & 1.06 & 4.5 & 3300 & $3.491 \pm 2.412$ &  $18.09^{+0.02}_{-0.02}$ &$-$3.29 \\
I19539+4424E & M5.5 & 1.03 & 5.0 & 2930 & $233 \pm 68^{\dagger}$ & $4.66^{+0.001}_{-0.001}$ &$-$2.57 \\
I19539+4424W & M5.5 & 1.02 & 5.0 & 3030 & $177 \pm 52^{\dagger}$ & $4.69^{+0.01}_{-0.01}$ &$-$2.00\\
I20198+2256$^\star$ & M3.0 & 1.00 & 5.0& 3400 &$34.080\pm 0.074$&$29.31^{+0.06}_{-0.06}$ &$-$3.27\\
\textbf{I20298+0941} & M5.0 & 0.96 & 4.5 & 2980 & $133.811 \pm1.386$& $7.47^{+0.07}_{-0.07}$ &$-$2.28\\
I20435+2407$^\star$ &  M2.5 & 0.91 & 5.0 &3460& $46.961\pm 0.045$& $21.28^{+0.02}_{-0.02}$ &$-$2.27\\
\textbf{I21160+2951E}$^\star$ & M4.0 & 0.95 & 5.0& 3260& $49.214\pm 0.243$ &$20.30^{+0.10}_{-0.10}$ &$-$3.78\\
I21376+0137 &  M4.5 & 1.01 & 5.0& 3150& $75 \pm 22^{\dagger}$& $13.33\pm 3.91$& $-$9.40\\
\textbf{I21521+0537} & M3.5 & 0.96 & 5.0 &3350& $58 \pm 17^{\dagger}$ & $17.24\pm 5.05$ & $-$3.97\\
\textbf{I22012+2818} &  M4.5 & 0.95 & 5.0& 3140 & $109.841\pm0.059$& $9.10^{+0.004}_{-0.004}$ &$-$4.37\\
I22403+0754 &  M0.5 & 0.96 & 4.5& 3640 &$24.240\pm 0.039$  &$41.20^{+0.06}_{-0.06}$ & +0.33\\
\textbf{I22468+4420} &  M4.0 & 0.98 & 5.0& 3270& $198.011\pm 0.038$& $5.04^{+0.0009}_{-0.0009}$ &$-$4.63\\
\textbf{I22518+3145} &  M3.5 & 0.93 & 5.0 &3400 &  $38\pm 11^{\dagger}$ & $26.31\pm 7.61$ &$-$3.26\\
\textbf{I23060+6355} &  M0.5 & 1.01 & 4.5 &3650& $41.539 \pm0.028$ &$24.05^{+0.01}_{-0.01}$ &$-$1.44\\
I23548+3831$^\star$ &  M4.0 & 0.97 & 5.0 &3250& $59.346\pm 0.052$ &$16.84^{+0.01}_{-0.01}$ &$-$3.74\\
\textbf{I23578+3837}$^\star$ &  M3.5 & 0.96 & 5.0 &3380 & $47.305\pm 1.498$ &$21.16^{+0.69}_{-0.65}$ & $-$3.47\\
I13352+3010 &  M0.0 & 1.02 & 4.5 &3660& $25.688\pm 0.057$ &$38.88^{+0.087}_{-0.086}$ & +0.55\\
I16591+2058  & M3.5 & 0.96 & 5.0 &3370 & $56.182\pm 0.572$ &$17.79^{+0.18}_{-0.18}$ &$-$2.18\\
I17006+0618 &  M1.0 & 0.94 & 4.5 &3570 & $33.381\pm 0.040$ & $29.93^{+0.03}_{-0.03}$ & +0.37\\
I20105+0632  & M4.0 & 0.94 & 5.0& 3270& $62.367\pm0.077$ &$16.02^{+0.02}_{-0.02}$ & $-$4.39\\
\textbf{I23340+0010} &  M2.5 & 1.08 & 5.0& 3500 &$71.374\pm 0.049$ &$14.00^{+0.009}_{-0.009}$ & +0.43\\
\cutinhead{``Maybe" Flares }    
I00570$+$4505 & M3.0 & 0.98 & 5.0 & 3380 & $57\pm17^{\dagger}$  & $17.54\pm 5.23$ & $+$0.45\\
I05342$+$1019N & M3.0 & 1.04 & 5.0 & 3400 & $45.17\pm0.08$ & $22.22 \pm 0.039 $ & $+$0.34\\
I07446$+$0333 & M4.5 & 1.01 & 5.0 & 3130 & $167\pm59$ & $5.98\pm 0.002$ & $-$4.51 \\ 
I09302+2630 & M3.5 & 0.97 & 5.0 & 3290 & $41.15\pm7$ & $24.09\pm 0.04$ & $-$1.11 \\    
I12142+0037 & M5.0 & 1.04 & 5.0 & 3040 & $115\pm34^{\dagger}$ & $8.69\pm2.57$ & $-$3.88\\
I12436$+$2506 & M3.5 & 0.96 & 5.0 & 3280 & $43.65\pm0.24$ & $22.93\pm5.76$ & $-$0.08\\
I13229$+$2428 & M4.0 & 1.07 & 4.5 & 3130 & $60\pm18^{\dagger}$ & $16.66\pm5.0$ & $+$0.26 \\
I19032$+$6359 & M3.5 & 0.90 & 5.0 & 3380 & $45\pm13^{\dagger}$ & $22.22\pm 6.41$ & $-$3.10 \\
\enddata
\tablenotetext{a}{Targets in bold are classified as flare stars in Simbad. Parallax values with a $^{\dagger}$ come from the photometric parallaxes reported in \citet{lepine13}.}
\end{deluxetable*}

\section{Results} \label{sec:results}

\begin{figure*}
\begin{center}
\includegraphics[width=\textwidth]{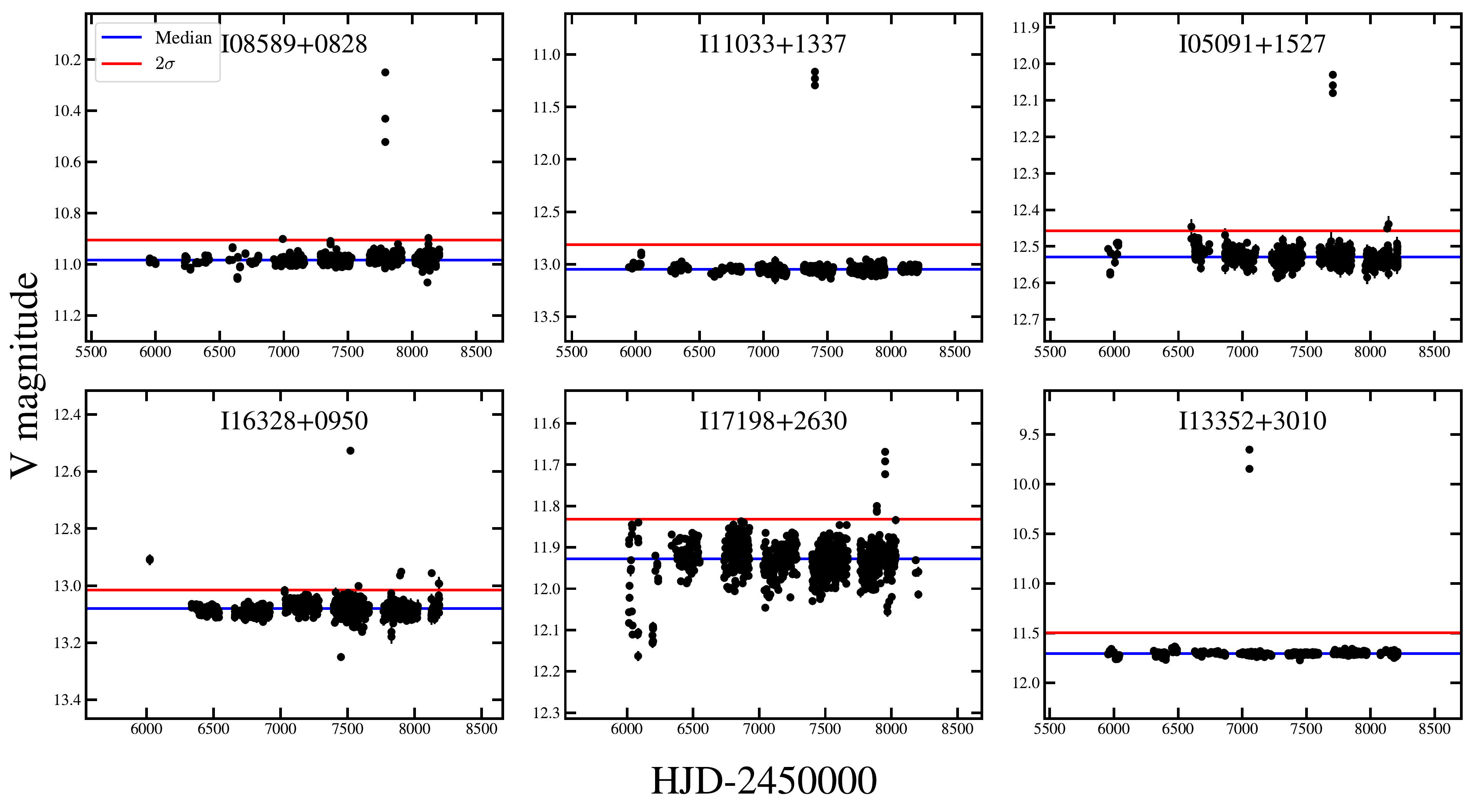}
\end{center}
\vspace{-3mm}
\caption{Representative examples of light curves from six of our flaring stars from the \citet{lepine13} sample. The red line shows the $2\sigma$ threshold for detection, while blue lines are the the mean magnitude. The top row consists of stars with H$\alpha$ in emission and the bottom row are stars with $\rm H\alpha$ in absorption or weak.}
\label{fig:fig2}
\end{figure*}

\begin{figure}
\begin{center}
\includegraphics[width=\columnwidth]{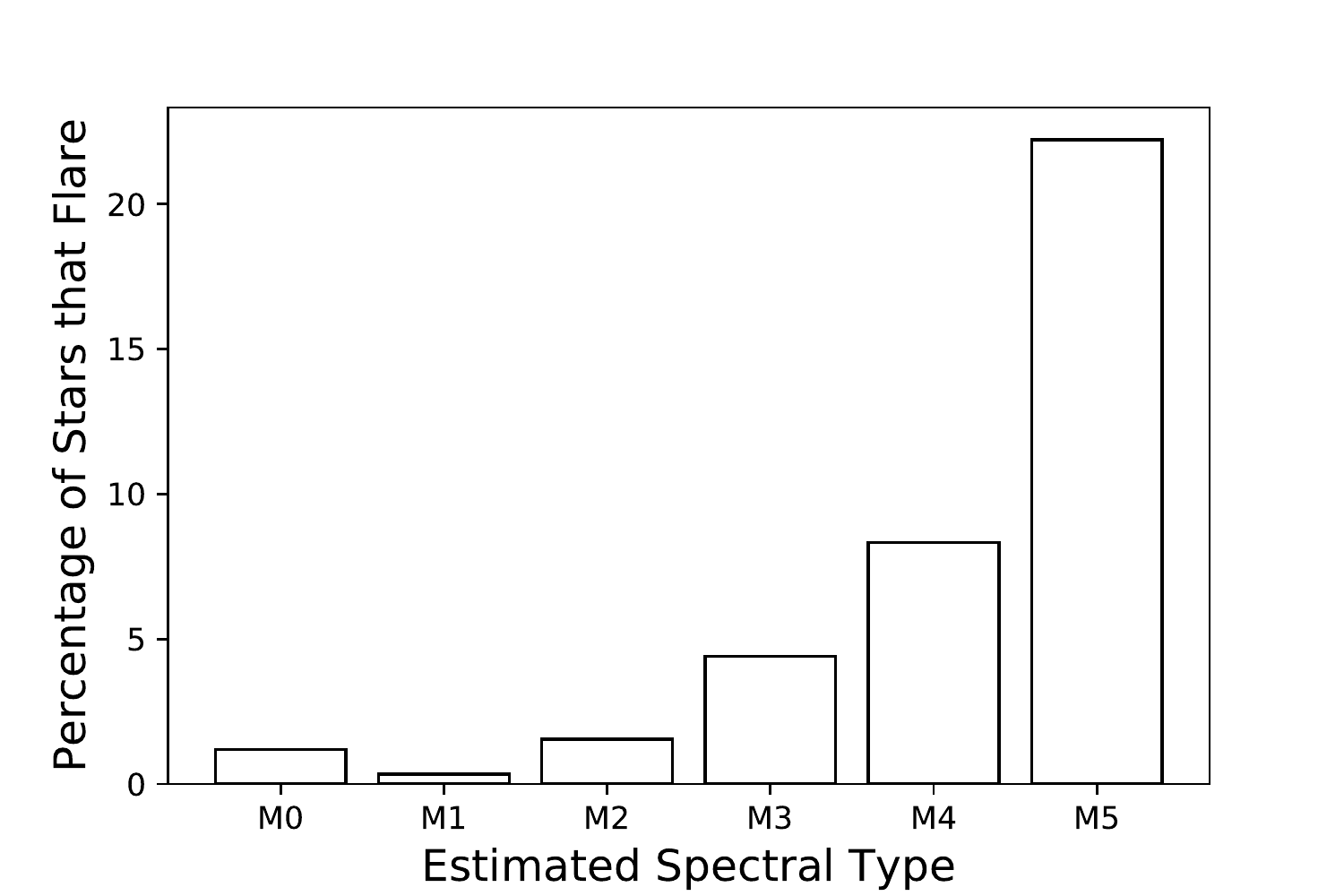}
\end{center}
\vspace{-3mm}
\caption{Distribution of spectral types among the final sample of flares from the \citet{lepine13} sample, representing $\sim$90\% of M dwarfs in the North. The histogram shows the number of stars from which we detected flares as a fraction of the total number of stars within each spectral class. The flaring fraction increases with spectral type, peaking at $\sim$25\% for the M5 class.}
\label{fig:fig3}
\end{figure} 

We identified 62 flares on 62 stars from the \citet{lepine13} sample. Figure~\ref{fig:fig2} shows a few representative light curves, and Table~\ref{tab:stellarprops} contains the properties of these stars. Of these, 34 were stars with previously observed flares, but the remaining 28 ($\sim$45\%) are new.  The 34 on known flare stars are 27\% of the 122 in the sample, which gives a very rough sense of the ``completeness".  In practice, flare energies follow a power-law distribution, so completeness is a complex combination of the duty cycle of the observations, individual flare rates, and our amplitude sensitivity, so this should only be interpreted qualitatively. 

The flaring candidates range in spectral types between early M0 to M5 with  temperatures of 2850 K to 3730 K. Their average  distance is 62 pc (as inferred from \citealp{Bailer-jones18}), and they have a median quiescent brightness of $V$ = 12.6 mag. These stars may have flared multiple times over the duration of the ASAS-SN observations ($\sim$6 years), and visual inspection of the light curves indeed reveals several small events that passed the statistical cut. The amplitudes of most of the events are small, only a fraction of a magnitude. Specifically, they range from $0.12<\Delta V< 2.04$ mag. We list the $\Delta V$ and the energies (Section ~\ref{sec:energies}) of the highest-amplitude events of each star, and include the number of potential flares in their light curves (events that passed the statistical threshold but were not confirmed) in Table~\ref{tab:flareprops}. 

\begin{deluxetable*}{llccrrrc}[ht]
\tablenum{4}
\tablecaption{Flare Properties \label{tab:flareprops}}
\tablehead{
\colhead{ID} & \colhead{Spt} & \colhead{Potential} & \colhead{$\Delta$V} & \colhead{$E_{\rm V}$} &\colhead{$E_{\rm U}$} & \colhead{$E_{\rm bol}$}&\colhead{H$\alpha$ EW} \vspace{-2mm} \\
 \colhead{} & \colhead{} & \colhead{Flares} & \colhead{} & \colhead{(erg)} & \colhead{(erg)} & \colhead{(erg)} & \colhead{(\AA)}}
\startdata
I00118$+$2259 &  M3.5 & 3 & 0.22 & (1.99$\pm$0.95)$\times10^{32}$ & (3.59$\pm$1.72)$\times10^{32}$ & (4.72$\pm$2.26)$\times10^{33}$ & $+$0.30 \\
I00162$+$1951W &  M4 & 1 & 0.16 & (2.28$\pm$0.75)$\times10^{34}$& (4.11$\pm$1.35)$\times10^{34}$ &(5.41$\pm$1.78)$\times10^{35}$& $-$4.01 \\
I00325$+$0729 &  M4 & 3 & 1.02 & (6.88$\pm$3.19)$\times10^{33}$ & (1.23$\pm$0.57)$\times10^{34}$& (1.63$\pm$0.75)$\times10^{35}$& $-$3.97 \\
I01033$+$6221 & M5 & 3 & 0.24 & (5.92$\pm$2.60)$\times 10^{31}$ & (1.06$\pm$0.46)$\times10^{32}$ & (1.40$\pm$0.61)$\times10^{33}$& $-$8.06 \\
I01593$+$5831  & M4 & 1 & 0.28 & (2.81$\pm$1.14)$\times10^{32}$ &(5.05$\pm$2.05)$\times10^{32}$ & (6.65$\pm$2.07)$\times10^{33}$&$-$5.00 \\
I02002$+$1303 & M5 & 2 & 0.87 & (2.67$\pm$0.26)$\times10^{31}$ &(4.82$\pm$0.47)$\times10^{31}$ & (6.34$\pm$0.62)$\times10^{32}$& $-$0.83 \\
I02088$+$4926  & M4 & 5 & 0.15 & (1.32$\pm$0.36)$\times10^{32}$ &(2.39$\pm$0.66)$\times10^{32}$ & (3.14$\pm$0.87)$\times10^{33}$& $-$3.76 \\
I02336$+$2455 & M4.5 & 6 & 0.20 & (6.88$\pm$1.10)$\times10^{35}$&(1.23$\pm$0.19)$\times10^{36}$ & (1.63$\pm$0.26)$\times10^{37}$& $-$3.06 \\
I03394$+$2458 & M3.5 & 6 & 0.17 & (4.79$\pm$1.53)$\times10^{31}$ &(8.63$\pm$2.70)$\times10^{31}$ & (1.13$\pm$0.36)$\times10^{33}$&
$-$0.65\\
I04284$+$1741 & M2 & 2 & 0.12 & (1.26$\pm$0.57)$\times10^{33}$ &(2.26$\pm$1.04)$\times10^{33}$ & (2.98$\pm$1.37)$\times10^{34}$&
$-$1.95 \\
I04480$+$1703 & M0.5 & 1 & 0.14 & (1.23$\pm$0.34)$\times10^{33}$ &(2.22$\pm$0.61)$\times10^{33}$ & (2.93$\pm$0.80)$\times10^{33}$&
+0.07 \\
I05019$+$0108 & M4.5 & 7 & 0.20 & (4.06$\pm$1.24)$\times10^{32}$ &(7.31$\pm$2.23)$\times10^{32}$ & (9.62$\pm$2.94)$\times10^{33}$&
$-$3.89 \\ 
I05062+0439 & M4 & 3 & 0.56 & (1.13$\pm$0.57)$\times10^{33}$ &(2.04$\pm$1.03)$\times10^{33}$ & (2.68$\pm$1.36)$\times10^{34}$&
$-$5.89 \\ 
I05091+1527 & M3 & 2 & 0.50 & (2.12$\pm$0.93)$\times10^{33}$ &(3.82$\pm$1.68)$\times10^{33}$ & (5.03$\pm$2.22)$\times10^{34}$&
$-$1.80 \\
I05337+0156 & M3 & 4 & 0.14 & (6.60$\pm$1.61)$\times10^{31}$ &(1.18$\pm$0.29)$\times10^{32}$ & (1.56$\pm$0.38)$\times10^{33}$&
$-$ 4.98 \\
I05547+1055 & M3 & 3 & 0.15 & (4.43$\pm$1.10)$\times10^{32}$ &(7.06$\pm$1.76)$\times10^{32}$ & (9.29$\pm$2.31)$\times10^{33}$&
$-$3.55\\ 
I07364+0704 & M5 & 3 & 0.60 & (1.51$\pm$0.32)$\times10^{31}$ &(2.72$\pm$0.58)$\times10^{31}$ & (3.58$\pm$0.76)$\times10^{32}$&
$-$3.55 \\
I07384+2400 & M3.5 & 2 & 0.64 & (7.63$\pm$3.12)$\times10^{32}$ &(1.37$\pm$0.56)$\times10^{33}$ & (1.80$\pm$0.74)$\times10^{34}$&
$-$4.08 \\
I07558+8323 & M4 & 7 & 0.19 & (6.27$\pm$2.95)$\times10^{31}$ &(1.12$\pm$0.53)$\times10^{32}$ & (1.48$\pm$0.69)$\times10^{33}$&
$-$3.27 \\
I08589+0828 & M4.5 & 1 & 0.72 & (5.73$\pm$2.73)$\times10^{32}$ &(1.03$\pm$0.49)$\times10^{33}$ & (1.35$\pm$0.64)$\times10^{34}$&
$-$2.74 \\
I09177+4612 & M2.5 & 4 & 0.23 & (1.00$\pm$0.48)$\times10^{33}$ &(1.80$\pm$0.86)$\times10^{33}$ & (2.37$\pm$1.13)$\times10^{34}$&
$-$2.89 \\
I10360+0507 & M4.0 & 4 & 0.30 & (2.82$\pm$1.42)$\times10^{30}$ &(5.08$\pm$2.56)$\times10^{30}$ & (6.69$\pm$3.37)$\times10^{31}$&
$-$2.80 \\
I10367+1521 & M4.5 & 2 & 0.80 & (4.45$\pm$2.00)$\times10^{32}$ &(8.02$\pm$3.60)$\times10^{32}$ & (1.05$\pm$0.47)$\times10^{34}$&
$-$4.46 \\
I11033+1337 & M4 & 1 & 1.87 & (2.79$\pm$0.87)$\times10^{33}$&(5.02$\pm$1.57)$\times10^{33}$ & (6.61$\pm$2.07)$\times10^{34}$&
$-$1.06 \\
I12189+1107 & M5 & 5 & 0.28 & (1.08$\pm$0.42)$\times10^{31}$ &(1.96$\pm$0.76)$\times10^{31}$ & (2.58$\pm$1.00)$\times10^{32}$&$-$3.53 \\
I12332+0901 & M5.5 & 5 & 0.29 & (2.00$\pm$0.87)$\times10^{30}$ &(3.61$\pm$1.56)$\times10^{30}$ & (4.75$\pm$2.06)$\times10^{31}$&$-$4.51 \\
I12485+4933 & M3.5 & 5 & 0.23 & (3.94$\pm$1.87)$\times10^{32}$ &(7.09$\pm$3.37)$\times10^{32}$ & (9.34$\pm$4.43)$\times10^{33}$&
$-$4.73 \\
I13007+1222 & M1.5 & 3 & 0.14 & (8.25$\pm$3.95)$\times10^{32}$ &(1.48$\pm$0.71)$\times10^{33}$ & (1.95$\pm$0.93)$\times10^{34}$&
$-$1.96 \\
I13317+2916 & M4.5 & 7 & 0.20 & (2.97$\pm$1.42)$\times10^{32}$ &(5.35$\pm$2.57)$\times10^{32}$ & (7.04$\pm$3.82)$\times10^{33}$&
$-$7.51 \\
I15126+4543 & M4 & 4 & 0.29 & (3.18$\pm$1.52)$\times10^{31}$ &(5.73$\pm$2.74)$\times10^{31}$ & (7.55$\pm$3.61)$\times10^{32}$&
$-$3.19 \\
I15218+2058 & M2 & 1 & 0.17 & (6.47$\pm$3.10)$\times10^{32}$ &(1.16$\pm$0.55)$\times10^{33}$ & (1.53$\pm$0.73)$\times10^{34}$&
$-$2.25 \\
I15238+5609  & M1 & 1 & 0.23 & (3.03$\pm$0.26)$\times10^{33}$ &(5.45$\pm$0.48)$\times10^{33}$ & (7.18$\pm$0.63)$\times10^{34}$&
$-$1.97 \\
I15555+3512  & M4.5 & 4 & 0.33 & (1.07$\pm$0.47)$\times10^{31}$ &(1.93$\pm$0.86)$\times10^{31}$ & (2.55$\pm$1.13)$\times10^{32}$&
$-$4.72 \\
I15557+6840 & M2.5 & 6 & 0.15 & (6.02$\pm$3.01)$\times10^{32}$ &(1.08$\pm$0.54)$\times10^{33}$ & (1.42$\pm$0.71)$\times10^{34}$&
$-$2.84 \\
I15581+4927 & M1 & 7 & 0.51 & (2.14$\pm$0.41)$\times10^{33}$ &(3.86$\pm$0.07)$\times10^{33}$ & (5.08$\pm$0.09)$\times10^{34}$&
$-$2.95 \\
I16328+0950 & M3.5 & 4 & 0.55 & (1.87$\pm$0.93)$\times10^{32}$ &(3.36$\pm$1.69)$\times10^{32}$ & (4.43$\pm$2.22)$\times10^{33}$&
 +0.31 \\
 I17198+2630 &  M3.5 & 2 & 0.26 &  (2.07$\pm$0.98)$\times10^{32}$ &(3.73$\pm$1.77)$\times10^{32}$ & (4.91$\pm$2.33)$\times10^{33}$&
$-$0.35 \\
I17338+1655 & M5.5 & 4 & 0.67 &  (1.89$\pm$7.02)$\times10^{31}$ & (3.40$\pm$1.26)$\times10^{32}$ & (4.48$\pm$1.66)$\times10^{33}$ &
$-$9.16 \\
I18022+6415  & M5 & 4 & 0.73 & (4.74$\pm$1.60)$\times10^{31}$ &(8.53$\pm$2.88)$\times10^{31}$ & (1.12$\pm$0.37)$\times10^{33}$&
$-$3.17 \\
I18358+8005 & M4 & 2 & 0.45 & (2.68$\pm$1.34)$\times10^{32}$ &(4.83$\pm$2.42)$\times10^{33}$ & (6.35$\pm$3.19)$\times10^{33}$&
 $-$2.32 \\
 I18427+1354 & M4.5 & 4 & 0.60 &  (4.92$\pm$1.24)$\times10^{31}$ &(8.86$\pm$2.23)$\times10^{31}$ & (1.16$\pm$0.29)$\times10^{33}$&
$-$2.26 \\
I19146+1919 & M3.5 & 1 & 0.40 &  (1.22$\pm$0.14)$\times10^{34}$ &(2.20$\pm$0.26)$\times10^{33}$ & (2.90$\pm$0.34)$\times10^{35}$&
$-$3.29 \\
I19539+4424E & M5.5 & 4 & 0.30 &  (5.70$\pm$0.09)$\times10^{30}$ &(5.92$\pm$0.10)$\times10^{30}$ & (7.80$\pm$0.13)$\times10^{31}$&
$-$2.57 \\
I19539+4424W & M5.5 & 6 & 0.34 &  (2.49$\pm$0.23)$\times10^{30}$  &(4.49$\pm$0.41)$\times10^{30}$ & (5.90$\pm$0.54)$\times10^{31}$&
$-$2.00 \\
\enddata
\tablecomments{``Potential flares" is the number of events that pass the 2$\sigma$ statistical cut, and $\Delta$V is the magnitude of the largest flare observed. The potential flares are not necessarily confirmed; in some cases, only the largest flare was confirmed. $E_{\rm U}$ and $E_{\rm bol}$ are estimated from $E_{\rm V}$ using the relations from \citet{lacy76} and \citet{gunther19}.}
\end{deluxetable*}

\begin{deluxetable*}{llcccccc}[ht]
\tablenum{4}
\tablecaption{Flare Properties (continued) \label{tab:flareprops}}
\tablehead{
\colhead{ID} & \colhead{Spt} & \colhead{Potential} & \colhead{$\Delta$V} & \colhead{$E_{\rm V}$} &\colhead{$E_{\rm U}$} & \colhead{$E_{\rm bol}$}&\colhead{H$\alpha$ EW} \vspace{-2mm} \\
 \colhead{} & \colhead{} & \colhead{Flares} & \colhead{} & \colhead{(erg)} & \colhead{(erg)} & \colhead{(erg)} & \colhead{(\AA)}}
\startdata
I20198+2256 & M3  & 1 & 0.23 & (2.82$\pm$0.58)$\times10^{32}$ &(5.08$\pm$1.05)$\times10^{32}$ & (6.68$\pm$1.38)$\times10^{33}$&
$-$3.27 \\
I20298+0941 & M5  & 4 & 0.78 &  (7.32$\pm$3.61)$\times10^{31}$ &(1.31$\pm$0.64)$\times10^{32}$ & (1.73$\pm$0.85)$\times10^{33}$&
$-$2.28 \\
I20435+2407 & M2.5 & 3 & 0.31 &  (5.75$\pm$0.59)$\times10^{32}$ &(1.03$\pm$0.10)$\times10^{33}$ & (1.36$\pm$0.14)$\times10^{34}$&
$-$2.27 \\
I21160+2951E & M4 & 8 & 0.12 &  (7.14$\pm$1.17)$\times10^{31}$ &(1.28$\pm$0.21)$\times10^{32}$ & (1.69$\pm$0.27)$\times10^{33}$&
$-$3.78 \\
I21376+0137 & M4.5 & 8 & 0.12 &  (1.55$\pm$0.60)$\times10^{33}$ &(2.94$\pm$1.15)$\times10^{33}$ & (3.87$\pm$1.52)$\times10^{34}$&
$-$9.40 \\
I21521+0537 & M3.5 & 3 & 0.23 &  (2.60$\pm$1.32)$\times10^{32}$ &(4.68$\pm$2.38)$\times10^{32}$ & (6.17$\pm$3.13)$\times10^{33}$&
$-$3.97 \\
I22012+2818 & M4.5 & 2 & 0.73 & (1.19$\pm$0.41)$\times10^{32}$ &(2.14$\pm$0.74)$\times10^{32}$ & (2.82$\pm$0.98)$\times10^{33}$&
$-$4.37 \\
I22403+0754 & M0.5 & 5 & 0.18 &  (4.10$\pm$1.81)$\times10^{33}$ &(7.38$\pm$3.26)$\times10^{33}$ & (9.71$\pm$4.29)$\times10^{34}$&
+0.33 \\
I22468+4420 & M4 & 5 & 0.18 &  (3.48$\pm$1.10)$\times10^{31}$ &(6.26$\pm$1.98)$\times10^{31}$ & (8.24$\pm$2.61)$\times10^{32}$&
$-$4.63 \\
I22518+3145 & M3.5 & 2 & 0.82 & (2.04$\pm$0.49)$\times10^{32}$ &(3.68$\pm$0.85)$\times10^{32}$ & (4.85$\pm$1.16)$\times10^{33}$&
$-$3.26 \\
I23060+6355 & M0.5 & 2 & 0.19 & (2.23$\pm$0.44)$\times10^{33}$ &(3.68$\pm$0.88)$\times10^{32}$ & (5.29$\pm$1.05)$\times10^{33}$&
$-$1.44 \\
I23548+3831 & M4 & 2 & 0.19 & (4.89$\pm$5.04)$\times10^{32}$ &(8.80$\pm$9.08)$\times10^{32}$ & (1.15$\pm$1.19)$\times10^{34}$&
$-$3.74 \\
I23578+3837 & M3.5 & 5 & 0.26 & (2.09$\pm$0.60)$\times10^{32}$ &(3.77$\pm$1.09)$\times10^{32}$ & (4.96$\pm$1.43)$\times10^{33}$&
$-$3.47 \\
I13352+3010  & M0 & 1 & 2.04 & (5.61$\pm$2.34)$\times10^{34}$ &(1.01$\pm$0.42)$\times10^{35}$ & (1.32$\pm$0.55)$\times10^{36}$&
$+$0.55 \\
I16591+2058  & M3.5 & 1 & 1.11 & (5.34$\pm$0.91)$\times10^{32}$ &(3.97$\pm$0.68)$\times10^{32}$ & (5.23$\pm$0.90)$\times10^{33}$&
$-$2.18\\
I17006+0618 & M1 & 1 & 0.45 & (2.43$\pm$1.23)$\times10^{33}$ &(4.37$\pm$2.22)$\times10^{33}$ & (5.75$\pm$2.93)$\times10^{34}$&
$+$0.37 \\
I20105+0632 & M4 & 4 & 0.12 & (4.62$\pm$1.50)$\times10^{31}$ &(8.31$\pm$2.70)$\times10^{31}$ & (1.09$\pm$3.55)$\times10^{32}$&
$-$4.39 \\ 
I23340+0010 & M2.5 & 3 & 0.27 & (3.94$\pm$1.09)$\times10^{32}$ &(7.09$\pm$1.96)$\times10^{32}$ & (9.33$\pm$2.59)$\times10^{33}$&
$+$0.43 
\enddata
\end{deluxetable*}

Figure~\ref{fig:fig3} shows a histogram of all the stars with confirmed flares binned by spectral type from M0 to M5. The fraction of stars by spectral type with confirmed flares rises from M1 to M5, peaking at 25\% for M5, consistent with previous studies of M-dwarf flares \citep{kowalski09,yang:2017,mondrik18,gunther19}. \citet{yang:2017} identified 540 M dwarfs with flares from \textit{Kepler} long-cadence data and discovered that the active fraction rises steeply near the M4 subtype, coinciding with the threshold at which M dwarfs become fully convective. \citet{gunther19} found a similar trend from a systematic study of 763 flaring M stars in \textit{TESS} ranging from M0 to M8, with flare activity peaking for the M5 stars. Additionally, \citet{west:2004} examined 8000 late-type dwarfs in the Sloan Digital Sky Survey using the H$\alpha$ emission line as an activity indicator, and they showed that the fraction of active stars peaks near M8, in agreement with past work (\citealp{hawley96,gizis00,kowalski09,schmidt16}). \citet{kowalski09} attributed this result to a combination of increased flare visibility (the contrast between flare emission and the quiescent background emission of the star; \citealp{gershberg72}) and an increase in the active fraction. They also proposed that later-type stars maintain their activity for longer: nearly 10 Gyr for M8 stars versus $<$ 1 Gyr for M0 stars \citep{shields16,west2006}. 

\begin{figure}
\vspace{.1in}
\centering
\includegraphics[width=1\linewidth]{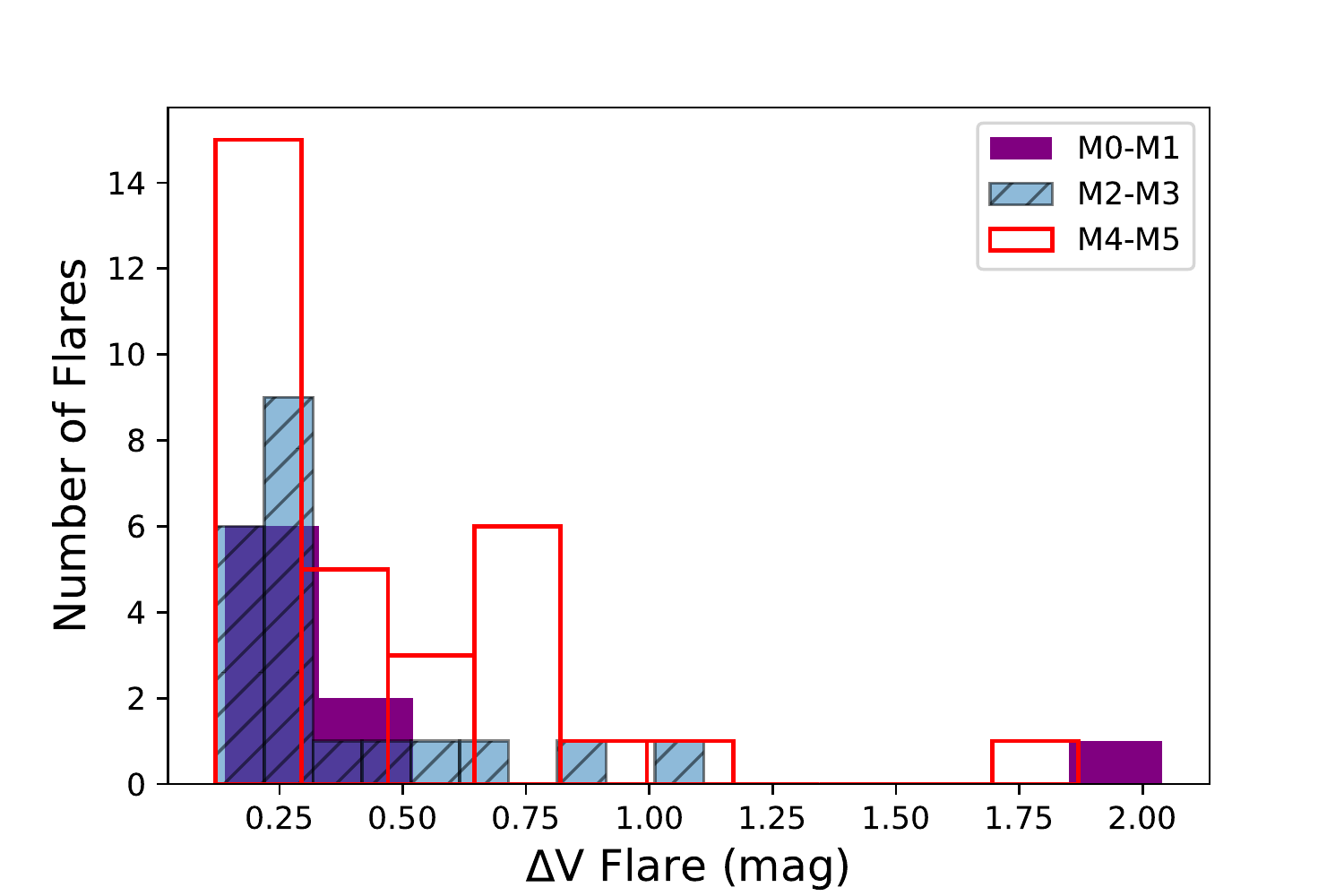}
%\vspace{-.25in}
\includegraphics[width=1\linewidth]{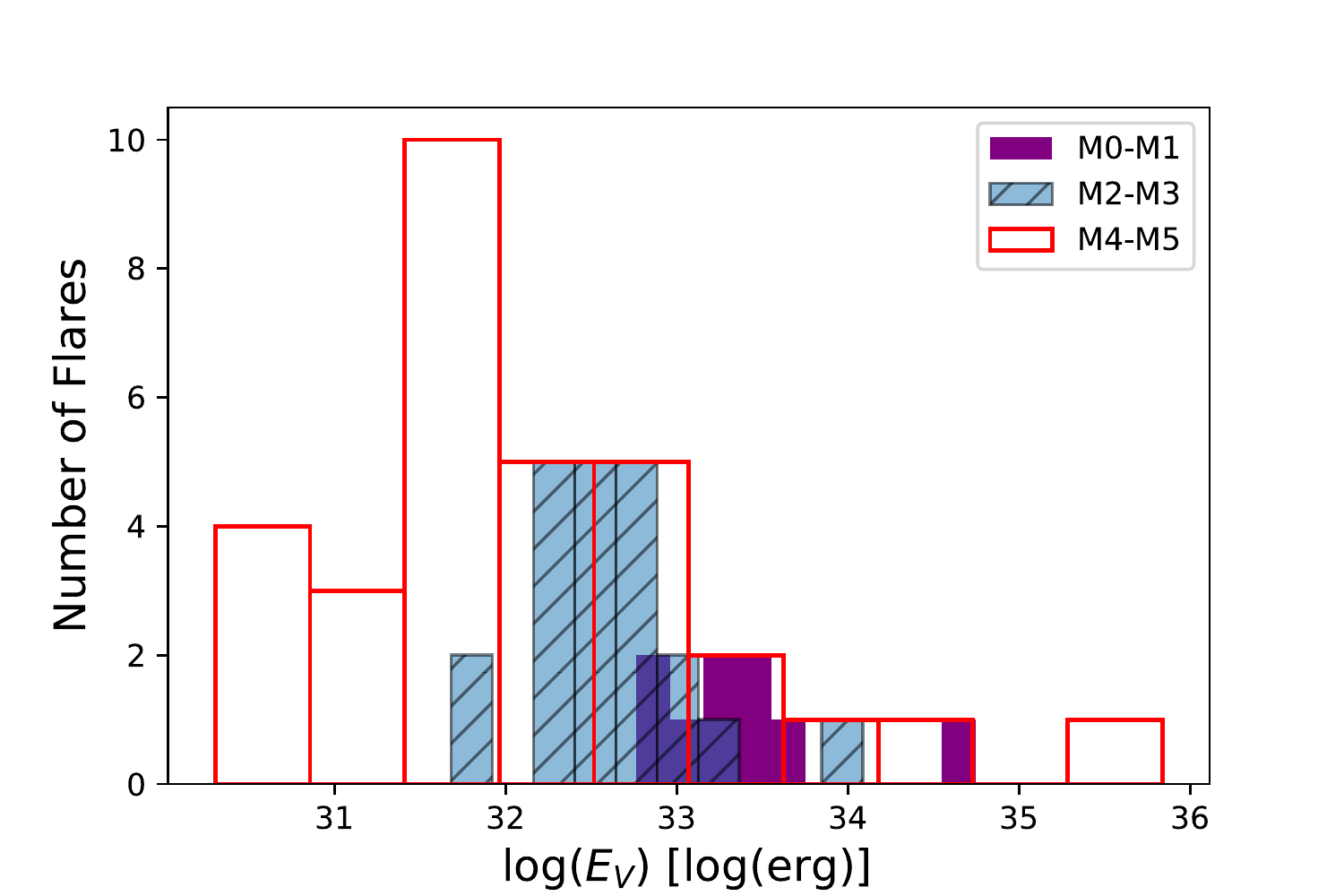}
\caption{Distribution of $\Delta$V mag (\textbf{top}) and V-band energies $E_{\rm V}$ (\textbf{bottom}) for the sample of validated flares divided into 3 spectral type groups: M0--M1, M2--M3, and M4--M5 stars.}
\label{fig:fig4} 
\end{figure}

In Figure~\ref{fig:fig4} and Table~\ref{table:statistics}, we show the distributions of $\Delta V$ and flare energies (see Section \ref{sec:energies}) broken down by spectral categories: M0--M1, M2--M3, and M4--M5. The M4--M5 group have the most flares among our sample, and they span a wider range of flare amplitudes than the M0--M1 and M2--M3 subgroups. Most of the M4--M5 flares have 0.25 $\lesssim \Delta V \lesssim $ 0.5 mag and jump to a $\Delta V \approx 2$ mag for I13352$+$3010, the largest amplitude confirmed flare. 

\begin{deluxetable}{llcc}
\tablenum{5}
\tablecaption{Flaring Statistics by Spectral Type \label{table:statistics}}
\tablehead{
\colhead{Spt} & \colhead{\# Stars} & \colhead{Median $\Delta$V (mag)} & \colhead{Median $E_{\rm V}$ (erg)}}
\startdata
M0--M1 & 9 &  0.19 & $2.2\times10^{33}$  \\ 
M2--M3 & 21 &  0.23 & $3.9\times10^{32}$ \\
M4--M5 & 32 & 0.30 & $7.2\times10^{31}$ \\
\enddata
\end{deluxetable}

All of the stars in the \citet{lepine13} sample have $\rm H\alpha$ equivalent width (EW) measurements as a proxy for activity \citep{walkowicz09}. \citet{lepine13} and \citet{west11} define active stars as those with $\rm H\alpha$ EWs $<$ $-0.75 \rm \AA$. The equivalent width of a line  measures the strength of the spectral feature relative to the continuum, where the continuum varies with spectral type. Therefore, rather than using the EWs directly from \citet{lepine13}, we use the $\chi$ ratio, defined as the ratio of the H$\alpha$ continuum to the bolometric luminosity \citep{west08}. Multiplied by the H$\alpha$ EW, it gives the ratio of the H$\alpha$ luminosity to the bolometric luminosity, and it allows a direct comparison between the strengths of the H$\alpha$ features for stars of different spectral types.

We observe a strong correlation between activity and flaring: most of the stars with flares have negative $\chi$ values (see Figure~\ref{fig:fig5}). However, some stars with flares detected by ASAS-SN have low H$\alpha$ emission (the bottom row of Figure~\ref{fig:fig2} shows flares of inactive stars). Of the 62 stars with flares, 55 stars are active ($\chi <0$) and 7 are inactive ($\chi >0$). 

This result is consistent with other studies of M-dwarf flares. In particular, \citet{hilton11}, \citet{kowalski09}, and \citet{hawley14} studied populations of active and inactive M dwarfs and detected flares among both types, although they also found that active stars flare more frequently than inactive ones. \citet{newton17} showed that rapidly rotating stars (which tend to be younger with ages $\leq$2~Gyr: \citealt{newton16}) exhibit higher H$\alpha$ emission than slower rotators (which tend to be older, with ages $\geq$5~Gyr).

Finally, although we were not able to confirm flares on M dwarfs with spectral types later than M5, we note that a total of 40 stars between M5 and M9 were automatically flagged by our algorithm in the \citet{lepine13} [18] and \citet{cruz07} [22] samples. It is likely that a fraction of those stars are truly flaring but did not meet our validation criteria. Despite the difficulty of characterizing flares on the lowest-mass stars, studies have found them to be common (e.g. \citealp{schmidt14, schmidt16, gizis17, vida17,paudel18,paudel19}).

\subsection{Flare energies}
\label{sec:energies}

The relatively coarse sampling of our data gives little information about the morphology of individual flares, which is important for calculating flare energies (see Figure 6 in \citealt{davenport14} for an example of a 1-minute cadence light curve of a classical and complex flare). The energy of a flare is defined as the product of its equivalent duration (ED) and the quiescent luminosity of the star. The equivalent duration is the integrated area under the light curve and is measured in seconds \citep{gershberg72}. We computed the quiescent luminosity of each star by multiplying the flux in the ASAS-SN bandpass by $4\pi d^{2}$, where $d$ is the distance to the source as derived from \citet{Bailer-jones18}, who uses the parallaxes from {\it Gaia} DR2 \citep{gaiadr2}. For sources undetected by {\it Gaia}, we used the photometric parallaxes reported in \citet{lepine13}. Although the limited cadence of our data prevents measurement of the ED of the flares directly, we can still place lower limits on the energies. We follow and briefly describe the methodology from \citet{schmidt18} to obtain flare EDs from a few detections. 

As in \citet{schmidt18}, we fit our data by modifying the classical flare model from \citet{davenport16}. They characterized a flare using two free parameters: a scaled amplitude and the full-time width at half the maximum flux, denoted $t_{1/2}$.  Their best-fit solution for the sharp rise is a fourth order polynomial of the form
\begin{eqnarray}
F_{\rm rise} = 1 + 1.941t_{1/2} - \nonumber 0.175t_{1/2}^{2} - 2.246t_{1/2}^{3}  - \nonumber \\ 1.125t_{1/2}^{4}
\end{eqnarray}

while the decay, which contains 61\% of the total energy, can be modeled as the sum of two exponentials:

\begin{eqnarray}
F_{\rm decay} = & 0.6890 & e^{-1.600t_{1/2}} + \nonumber \\
& 0.3030 & e^{-0.2783t_{1/2}} 
\end{eqnarray}

\begin{figure}
\begin{center}
\includegraphics[width=\columnwidth]{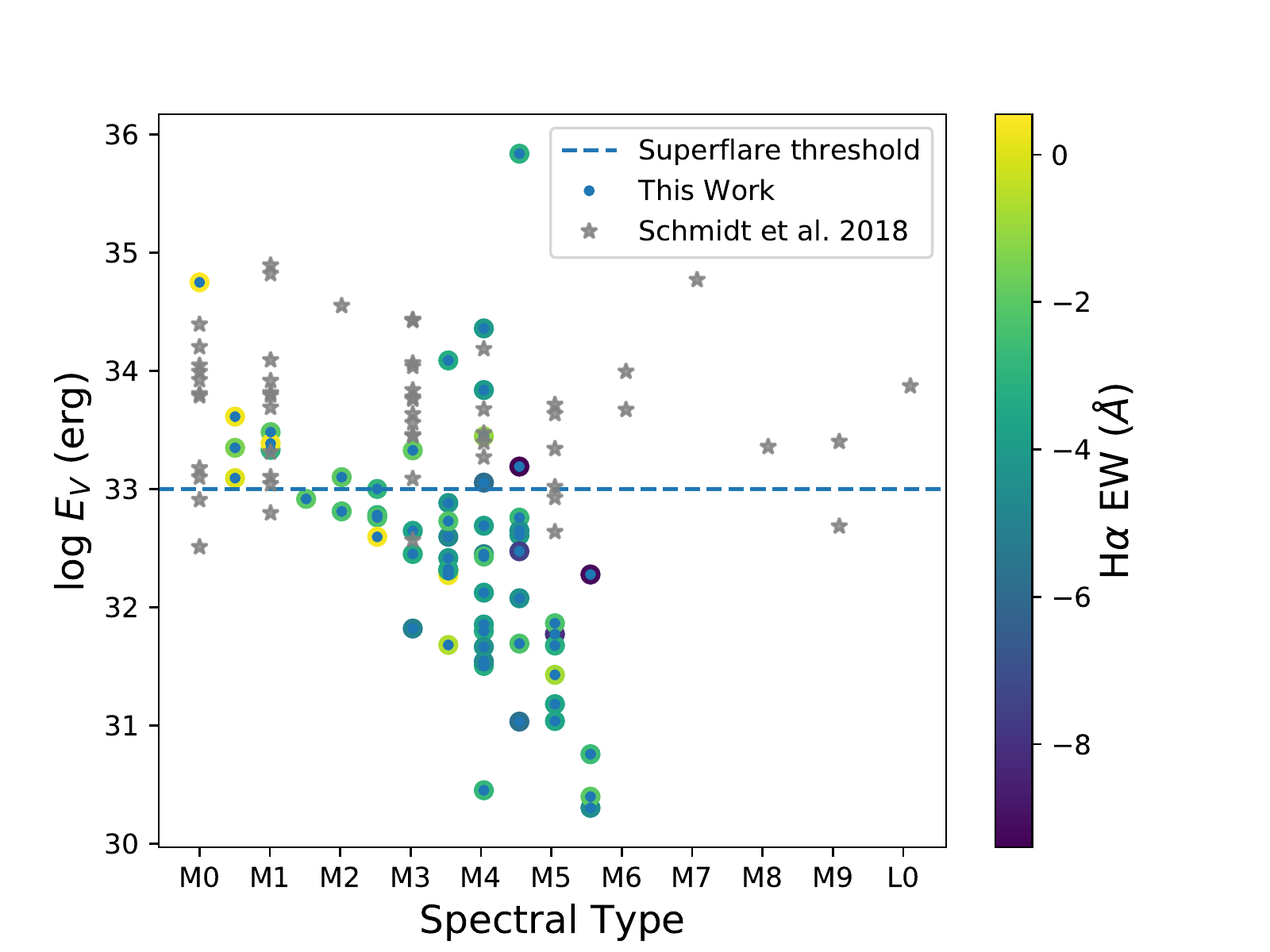}
\end{center}
%\vspace{-3mm}
\caption{Energy in V-band ($E_{\rm V}$) of the largest confirmed flare for each star by spectral subtype. The ``superflare" threshold marked by the horizontal dashed line denotes flares with V-band energies $\ga 10^{33}$ erg. The colors on the data points represent the values of the $\chi$ ratio, which is proportional to the ratio of the H$\alpha$ luminosity to the bolometric luminosity of a star.}
\label{fig:fig5}
\end{figure}

We estimated the flare energies by generating 4000 flare light curves varying the position of the peak and $t_{1/2}$ ($10 <t_{1/2}< 2000$ seconds). We then integrated the flux of the flare compared to the stellar luminosity for each combination of peak position and $t_{1/2}$ and took the median of all the values as the estimate of the equivalent duration. After applying this procedure to our sample, we derived energies ranging from $2.0\times10^{30}$ $\rm <E_{V}<$ $6.9\times10^{35}$ ergs. The energies are listed in Table~\ref{tab:flareprops}, and we plot the flare energies as a function of spectral type in Figure~\ref{fig:fig5}.

To estimate the bolometric energy $E_{\rm bol}$ of the flare from the $V$-band energy $E_{\rm V}$, we used the conversions from \citet{lacy76} and in \citet{gunther19}. \citet{lacy76} determined flare energy scaling relations based on simultaneous, multi-wavelength observations of flares in the $U$, $B$ and $V$ filters. They found that the $U$, $B$ and $V$ energies scale as $E_{\rm U} = 1.2 E_{\rm B}$ and $E_{\rm U} = 1.8 E_{V}$. 
Based on M dwarfs observed with \textit{TESS}, \citet{gunther19} convert their derived bolometric flare energies to $U$-band energies as $E_{\rm U} \approx 0.076 E_{\rm bol}$. Combining these relations, we find that $E_{\rm V} \approx 0.042 E_{\rm bol}$, and we list the resulting $E_{\rm bol}$ in Table~\ref{tab:flareprops}.

\subsection{Variability Analysis}
\label{sec:rotation}

To explore the connection between magnetic activity and stellar rotation, we estimated rotational periods for the stars in the \citet{lepine13} sample from the ASAS-SN light curves. We did not consider the \citealt{cruz07} sample here because of the stars' faintness. We used the \verb"astropy" implementation of the Generalized Lomb-Scargle (GLS; \citealt{2009A&A...496..577Z,1982ApJ...263..835S}) periodogram and the \verb"astrobase" implementation \citep{2018zndo...1246769B} of the Box Least Squares (BLS; \citealt{2002A&A...391..369K}) periodogram to search for periodicity over the range $0.05\leq P \leq100$ days. Periodic sources were classified using a random forest classifier and quality checks as described in \citet{2019MNRAS.486.1907J}. In order to minimize false positives due to spurious variability signals in the data, we implemented cuts in classification probability $\rm Prob>0.9$ and the Lafler-Kinmann string length statistic \citep{1965ApJS...11..216L,2002A&A...386..763C,2019MNRAS.486.1907J} calculated for the magnitudes sorted by phase $T(\phi|P)<0.65$. 

Among the 1276 dwarfs in the  \citet{lepine13} sample, we identified 77 rotational variables (see Table~\ref{tab:rotvariablestars}) and 40 other variables (see Table~\ref{tab:genvariablestars}). The source I16343$+$5709 had $\rm Prob=0.80$, but upon visual verification, we found that it was the known detached eclipsing binary CM Draconis with $\rm P=1.2684$ days \citep{morales09}. 

\begin{deluxetable*}{lrr| lrr}[t]
\tablenum{6} 
\tablecaption{Rotational Variable Stars \label{tab:rotvariablestars}}
\tablehead{
\colhead{ID} & \colhead{Probability} & \colhead{Period (days)} & \colhead{ID} & \colhead{Probability} & \colhead{Period (days)}}
\startdata
I00505$+$2449S & 0.991 & 9.656 & I09275$+$5039  & 0.978 & 18.93 \\
I00081$+$4757 & 0.963 & 8.751 & I09352$+$6114   & 0.917 & 97.32 \\
I00084$+$1725   & 0.979 & 15.363 & I10043$+$5023   & 0.975 & 3.067 \\
I00268$+$7008   & 0.974 & 18.195 & I10367$+$1521$^*$   & 0.946 & 7.153 \\
I01036$+$4051   & 0.998 & 28.445 & I11118$+$3332S  & 0.968 & 15.54 \\
I01066$+$1516   & 0.973 & 35.010 & I11519$+$0731   & 0.995 & 6.878 \\
I02028$+$0447   & 0.995 & 15.869 & I12156$+$5239   & 0.965 & 4.891 \\
I02180$+$5616   & 0.982 & 1.782 & I12485$+$4933$^*$   & 0.993 & 4.063 \\
I02292$+$1932   & 0.999 & 5.546 & I13518$+$1247   & 0.957 & 4.967 \\
I03168$+$8205   & 0.976 & 16.640  & I14023$+$1341   & 0.974 & 1.496 \\
I03247$+$4447   & 0.972 & 2.053 & I14153$+$1523   & 0.986 & 23.93  \\
I03322$+$4914S  & 0.999 & 5.947 & I15280$+$2547   & 0.984 & 26.17 \\
I03466$+$8207   & 0.977 & 1.248 & I15557$+$6840$^*$  & 0.994 & 3.920 \\
I04274$+$2022   & 0.982 & 43.61 & I15581$+$4927$^*$   & 0.977 & 4.492 \\
I04312$+$4217   & 0.999 & 3.960 & I15597$+$4403   & 0.991 & 1.480 \\
I04350$+$0839   & 0.999 & 11.75 & I16046$+$2620 & 0.973 & 16.80   \\
I04473$+$0627   & 0.987 & 87.25  & I16139$+$3346   & 0.982 & 24.59 \\
I04595$+$0147   & 0.999 & 4.384 & I16220$+$2250   & 0.994 & 5.489 \\
I00483$+$7116   & 0.990  & 9.488 & I17225$+$0531   & 0.964 & 67.65  \\
I05019$+$0108$^*$   & 0.991 & 4.187 & I17338$+$1655$^*$   & 0.919 & 1.325 \\
I05062$+$0439$^*$   & 0.962 & 2.666 & I18134$+$0526   & 0.944 & 0.793 \\
I05228$+$2016   & 0.938 & 28.93 & I18312$+$0650   & 0.940  & 1.046 \\
I05334$+$4809   & 0.977 & 14.27 & I18519$+$1300 & 0.994 & 5.971 \\
I05341$+$4732   & 0.999 & 12.428 & I18554$+$0824   & 0.992 & 12.58 \\
I05365$+$1119   & 0.988 & 2.018 & I19026$+$3231   & 0.995 & 3.545  \\
I05402$+$1239   & 0.960  & 6.331 & I19270$+$1242   & 0.986 & 3.478  \\
I05547$+$1055$^*$   & 0.987 & 1.130 & I19354$+$3746   & 0.994 & 8.258 \\
I06025$+$3707   & 0.975 & 31.53 & I20198$+$2256$^*$   & 0.944 & 97.16  \\
I06147$+$4727   & 0.977 & 27.20 & I20220$+$2147   & 0.974 & 38.20   \\
I06212$+$4414   & 0.991 & 5.728 & I20435$+$2407$^*$   & 0.987 & 2.398  \\
I06237$+$0502   & 0.967 & 71.35  & I21152$+$2547   & 0.988 & 48.60   \\
I06262$+$2349   & 0.967 & 15.77 & I21160$+$2951E$^*$  & 0.985 & 1.759 \\
I06310$+$5002   & 0.995 & 15.25 & I22006$+$2715   & 0.999 & 10.59 \\
I07295$+$3556   & 0.989 & 2.673 & I22107$+$0754   & 0.980  & 14.91 \\
I07346$+$2220   & 0.980  & 27.94 & I22129$+$5504N  & 0.966 & 4.834 \\
I07349$+$1445   & 0.982 & 20.08 & I23318$+$1956E  & 0.958 & 15.28 \\
I07384$+$2400$^*$   & 0.972 & 3.857 & I23548$+$3831$^*$   & 0.965 & 14.29 \\
I08317$+$0545   & 0.979 & 1.786 & I23578$+$3837$^*$   & 0.952 & 2.665 \\
I09161$+$0153   & 0.972 & 4.308 & \\
\enddata
\tablecomments{Objects with a $^{*}$ are those identified to have flares in the ASAS-SN data.}
\end{deluxetable*}

14 rotational variables had flares identified in the ASAS-SN data, which amounts to ${\sim}23\%$ of the flaring sample. 13 of those have rotation periods between $1.1 < P < 14.2 $ days (and an additional one has $P = 97.1$ d) and a median value of $P =$ 3.8 days. They have H$\alpha$ EWs of $< -2$\AA~and a median EW of $-$3.7\AA. On the other hand, the rotational variables with no confirmed flares have weaker H$\alpha$ emission, with EWs of $-$4.4\AA~to $+$0.62\AA~and a median value of $-0.24$\AA. Thus, the rotational variables with identified flares are more active than those without detected flares.  In agreement with previous studies (e.g., \citealt{newton17,doyle18}), we find that more ``active" stars (which exhibit starspot modulations and flares) have rotation periods of $\la$10 days, while the more inactive stars (as defined by weaker H$\alpha$ emission and lack of detectable flares) have longer rotation periods. There are exceptions to this trend, such as the active, flaring star I20198$+$2256 (which has $P\sim$ 97 days) or the active M5.5 star Proxima Centauri (which has $P\sim$ 83 days), both slow rotators compared to other stars with comparable flare rates and activity levels \citep{davenport16}.

\begin{deluxetable*}{lrr| lrr}[t]
\tablenum{7} 
\tablecaption{Other Variable Stars \label{tab:genvariablestars}}
\tablehead{
\colhead{ID} & \colhead{Probability} & \colhead{Period (days)} & \colhead{ID} & \colhead{Probability} & \colhead{Period (days)}}
\startdata
I00173$+$2910  & 0.967 & 39.45 & I13293$+$1126    & 0.927 & 57.13  \\
I01025$+$7140    & 0.924 & 50.90 & I15043$+$6023    & 0.951 & 38.43 \\
I01453$+$4632    & 0.964 & 4.093 & I17198$+$4142    & 0.952 & 62.24  \\
I01593$+$5831    & 0.961 & 4.144 & I18319$+$4041    & 0.966 & 45.47  \\
I02070$+$4938    & 0.978 & 28.00 & I20151$+$6331    & 0.976 & 28.94 \\
I02088$+$4926    & 0.965 & 1.496 & I20450$+$4429    & 0.931 & 75.75  \\
I02116$+$1833    & 0.946 & 61.93 & I21013$+$3314    & 0.965 & 2.960 \\
I03479$+$0247    & 0.965 & 3.990 & I21402$+$3703    & 0.983 & 2.990 \\
I04520$+$0628    & 0.908 & 50.66 & I21442$+$0638    & 0.987 & 62.68  \\
I05333$+$4448    & 0.958 & 42.16 & I21448$+$4417    & 0.923 & 78.61  \\
I06579$+$6219    & 0.918 & 53.88 & I21521$+$0537    & 0.985 & 0.938  \\
I07446$+$0333    & 0.951 & 11.11 & I22234$+$3227    & 0.985 & 5.937 \\
I08095$+$2154    & 0.971 & 25.46 & I22468$+$4420    & 0.943 & 4.362 \\
I08371$+$1507    & 0.940  & 53.74  & I22518$+$3145    & 0.950  & 4.922 \\
I09187$+$2645    & 0.970  & 26.96  & I23045$+$6645    & 0.954 & 26.39 \\
I09193$+$6203    & 0.919 & 2.922 & I23182$+$4617    & 0.962 & 17.38 \\
I09428$+$7002    & 0.969 & 24.10 & I23216$+$1717    & 0.936 & 78.22  \\
I10360$+$0507    & 0.929 & 6.843 & I23293$+$4128    & 0.963 & 4.252 \\
I11237$+$0833    & 0.969 & 39.69 & I23318$+$1956Wn  & 0.957 & 2.134  \\
I12428$+$4153    & 0.942 & 66.67 & I23431$+$3632    & 0.934 & 71.73  \\
\enddata
%\tablecomments{}
\end{deluxetable*}

\subsection{Influence on exoplanets}
\label{sec:habitability}

To find how many stars in our sample host confirmed exoplanets, we cross-matched our targets to the NASA Exoplanet Archive\footnote{https://exoplanetarchive.ipac.caltech.edu/} by coordinate and checked for planets found by any method. In the flaring sample, we found 1 confirmed exoplanet around I13007$+$1222 ($\Delta V = 0.14$, Spt = M1.5V) and one star with a threshold-crossing event (TCE), I19539$+$4424W, where a TCE is a series of transit-like features resembling a true exoplanet signature \citep{jenkins2002}. However, this star is in a binary with I19539$+$4424E, and therefore the TCE could be due to the stars eclipsing one another rather than a real planet.

In the full \citet{lepine13} sample, we found 6 confirmed planets (1 per star) associated with I04520$+$0628, I08551$+$0132, I11421$+$2642, I13007$+$1222, I16167$+$6714, and I16581$+$2544. Of those, I04520$+$0628 and I11421$+$2642 were identified as potential candidates by our flare-finding algorithm. We confirmed at least one flare around I13007$+$1222 with an energy of ($8.25 \pm 3.95$)$\times10^{32}$ erg.   

Flares can have adverse effects on the habitability of M-dwarf planets.  Long-term, frequent exposure to flares, coronal mass ejections, stellar proton events, and other related phenomena can cause atmospheric erosion over time (\citealp{lammer07,segura10,luger15,tilley17,howard18,howard:2019,loyd18}). For example, \citet{tilley17} modeled the impact of M-dwarf flares on an Earth-like, unmagnetized exoplanet in the habitable zone of its host star. They found that flares with energies comparable to $10^{34}$ erg in the $U$-band with a frequency $\ga$ 1 per month will erode $\sim$99$\%$ of the ozone layer. Converting $V$-band to $U$-band energies using $E_{\rm U} = 1.8 E_{V}$, we find 5 sources with flare energies $\ga 10^{34}$ erg. We did not construct a flare frequency distribution (FFD) because we only confirmed the largest flare in each star, and therefore we do not constrain the frequency of flares at this energy. However, we estimate that such high-energy events are relatively infrequent: in stars where we confirm a large flare, we do not find other flares of comparable amplitude (or energy) in the 6-year light curves. Flare distributions are power laws \citep{hawley14}, and high-energy flares are less common than low-energy ones.  We do see frequent, small flare-like features (some are apparent in the light curves in Figure~\ref{fig:fig2}), and  it is possible that even small, low-energy events like “microflares” ($\rm E_{V} \sim 10^{29}$ erg) that occur sufficiently frequently could cause irreversible damage to an Earth-like atmosphere and complex surface life (e.g., \citealp{gunther19}). 

\section{Conclusions} \label{sec:conclusions}

In this paper, we performed an optical search for stellar flares of $\sim$1400 M dwarfs using long-baseline, moderate-cadence photometry from ASAS-SN. The sample was comprised of the brightest ($J <9$ mag) M dwarfs in the northern hemisphere and the nearest ($d<20$ pc) M dwarfs. Applying a simple flare-finding algorithm, we automatically detected 480 stars with potential flares, of which 62 were confirmed after visual inspection of the data. The confirmed events range in amplitude of 0.12 $< \Delta V <$ 2.04 mag and have $V$-band energies of $2.0\times10^{30} \lesssim E_{V} \lesssim 6.9\times10^{35}$ erg, consistent with previous results in the literature. 

From our sample of confirmed flares, we find a strong trend in the flaring fraction as a function of spectral type: the cooler M dwarfs flare more, with 25\% of all M5 dwarfs in our sample flaring. We were not able to confirm flares among the later-types (from M6--M9) because the majority of the coolest dwarfs are optically faint and are therefore near or beyond the detection limits of the instruments. We find a positive correlation between the $\rm H\alpha$ equivalent width and the flaring fraction: stars in our confirmed flare sample display strong $\rm H\alpha$ emission. Finally, we derive rotation periods for stars in the \citet{lepine13} sample and find that the rotational variables with detected flares have stronger H$\alpha$ emission and relatively shorter periods than those without detected flares.

Flares and stellar activity impact the habitability of exoplanets around M dwarfs. Observations at higher cadence and across wavebands, such as with {\it TESS} and {\it JWST}, are crucial to measure flaring frequency and reliable flare bolometric energies. Overall, these constraints are crucial to probe the relationship between flares, magnetic fields, rotation, and stellar age.

\acknowledgements

We thank Dr. Jennifer van Saders and Dr. Kris Stanek for useful discussions that enriched this project. B.J.S. is supported by NSF grant AST-1908952.
ASAS-SN is supported by the Gordon and Betty Moore
Foundation through grant GBMF5490 to the Ohio State
University and NSF grant AST-1515927. Development of
ASAS-SN has been supported by NSF grant AST-0908816,
the Mt. Cuba Astronomical Foundation, the Center for Cosmology and AstroParticle Physics at the Ohio State University, the Chinese Academy of Sciences South America Center for Astronomy (CASSACA), the Villum Foundation, and
George Skestos.

%\appendix

\nocite{*}
\bibliographystyle{aasjournal}
\bibliography{cr_smc}

\begin{thebibliography}{}
\expandafter\ifx\csname natexlab\endcsname\relax\def\natexlab#1{#1}\fi
\providecommand{\url}[1]{\href{#1}{#1}}
\providecommand{\dodoi}[1]{doi:~\href{http://doi.org/#1}{\nolinkurl{#1}}}
\providecommand{\doeprint}[1]{\href{http://ascl.net/#1}{\nolinkurl{http://ascl.net/#1}}}
\providecommand{\doarXiv}[1]{\href{https://arxiv.org/abs/#1}{\nolinkurl{https://arxiv.org/abs/#1}}}

\bibitem[{{Allard} {et~al.}(2011){Allard}, {Homeier}, \& {Freytag}}]{allard10}
{Allard}, F., {Homeier}, D., \& {Freytag}, B. 2011, in Astronomical Society of
  the Pacific Conference Series, Vol. 448, 16th Cambridge Workshop on Cool
  Stars, Stellar Systems, and the Sun, ed. C.~{Johns-Krull}, M.~K. {Browning},
  \& A.~A. {West}, 91

\bibitem[{{Bailer-Jones} {et~al.}(2018){Bailer-Jones}, {Rybizki}, {Fouesneau},
  {Mantelet}, \& {Andrae}}]{Bailer-jones18}
{Bailer-Jones}, C.~A.~L., {Rybizki}, J., {Fouesneau}, M., {Mantelet}, G., \&
  {Andrae}, R. 2018, \aj, 156, 58, \dodoi{10.3847/1538-3881/aacb21}

\bibitem[{{Bhatti} {et~al.}(2018){Bhatti}, {lgbouma}, {Joshua}, \&
  {Price-Whelan}}]{2018zndo...1246769B}
{Bhatti}, W., {lgbouma}, {Joshua}, \& {Price-Whelan}, A. 2018,
  {Waqasbhatti/Astrobase: Astrobase V0.3.14}, v0.3.14,  Zenodo,
  \dodoi{10.5281/zenodo.1246769}

\bibitem[{{Bochanski} {et~al.}(2010){Bochanski}, {Hawley}, {Covey}, {West},
  {Reid}, {Golimowski}, \& {Ivezi{\'c}}}]{bochanski10}
{Bochanski}, J.~J., {Hawley}, S.~L., {Covey}, K.~R., {et~al.} 2010, \aj, 139,
  2679, \dodoi{10.1088/0004-6256/139/6/2679}

\bibitem[{{Borucki} {et~al.}(2010){Borucki}, {Koch}, {Basri}, {Batalha},
  {Brown}, {Caldwell}, {Caldwell}, {Christensen-Dalsgaard}, {Cochran},
  {DeVore}, {Dunham}, {Dupree}, {Gautier}, {Geary}, {Gilliland}, {Gould},
  {Howell}, {Jenkins}, {Kondo}, {Latham}, {Marcy}, {Meibom}, {Kjeldsen},
  {Lissauer}, {Monet}, {Morrison}, {Sasselov}, {Tarter}, {Boss}, {Brownlee},
  {Owen}, {Buzasi}, {Charbonneau}, {Doyle}, {Fortney}, {Ford}, {Holman},
  {Seager}, {Steffen}, {Welsh}, {Rowe}, {Anderson}, {Buchhave}, {Ciardi},
  {Walkowicz}, {Sherry}, {Horch}, {Isaacson}, {Everett}, {Fischer}, {Torres},
  {Johnson}, {Endl}, {MacQueen}, {Bryson}, {Dotson}, {Haas}, {Kolodziejczak},
  {Van Cleve}, {Chandrasekaran}, {Twicken}, {Quintana}, {Clarke}, {Allen},
  {Li}, {Wu}, {Tenenbaum}, {Verner}, {Bruhweiler}, {Barnes}, \&
  {Prsa}}]{borucki10}
{Borucki}, W.~J., {Koch}, D., {Basri}, G., {et~al.} 2010, Science, 327, 977,
  \dodoi{10.1126/science.1185402}

\bibitem[{{Carrington}(1859)}]{carrington1859}
{Carrington}, R.~C. 1859, \mnras, 20, 13, \dodoi{10.1093/mnras/20.1.13}

\bibitem[{{Chabrier} \& {Baraffe}(1997)}]{chabrier97}
{Chabrier}, G., \& {Baraffe}, I. 1997, \aap, 327, 1039

\bibitem[{{Clarke}(2002)}]{2002A&A...386..763C}
{Clarke}, D. 2002, \aap, 386, 763, \dodoi{10.1051/0004-6361:20020258}

\bibitem[{{Cruz} {et~al.}(2007){Cruz}, {Reid}, {Kirkpatrick}, {Burgasser},
  {Liebert}, {Solomon}, {Schmidt}, {Allen}, {Hawley}, \& {Covey}}]{cruz07}
{Cruz}, K.~L., {Reid}, I.~N., {Kirkpatrick}, J.~D., {et~al.} 2007, \aj, 133,
  439, \dodoi{10.1086/510132}

\bibitem[{{Davenport}(2016)}]{davenport16}
{Davenport}, J.~R.~A. 2016, \apj, 829, 23, \dodoi{10.3847/0004-637X/829/1/23}

\bibitem[{{Davenport} {et~al.}(2012){Davenport}, {Becker}, {Kowalski},
  {Hawley}, {Schmidt}, {Hilton}, {Sesar}, \& {Cutri}}]{davenport12}
{Davenport}, J.~R.~A., {Becker}, A.~C., {Kowalski}, A.~F., {et~al.} 2012, \apj,
  748, 58, \dodoi{10.1088/0004-637X/748/1/58}

\bibitem[{{Davenport} {et~al.}(2014){Davenport}, {Hawley}, {Hebb},
  {Wisniewski}, {Kowalski}, {Johnson}, {Malatesta}, {Peraza}, {Keil},
  {Silverberg}, {Jansen}, {Scheffler}, {Berdis}, {Larsen}, \&
  {Hilton}}]{davenport14}
{Davenport}, J.~R.~A., {Hawley}, S.~L., {Hebb}, L., {et~al.} 2014, \apj, 797,
  122, \dodoi{10.1088/0004-637X/797/2/122}

\bibitem[{{Doyle} {et~al.}(2018){Doyle}, {Ramsay}, {Doyle}, {Wu}, \&
  {Scullion}}]{doyle18}
{Doyle}, L., {Ramsay}, G., {Doyle}, J.~G., {Wu}, K., \& {Scullion}, E. 2018,
  \mnras, 480, 2153, \dodoi{10.1093/mnras/sty1963}

\bibitem[{{Dressing} \& {Charbonneau}(2015)}]{dressing15}
{Dressing}, C.~D., \& {Charbonneau}, D. 2015, \apj, 807, 45,
  \dodoi{10.1088/0004-637X/807/1/45}

\bibitem[{{France} {et~al.}(2013){France}, {Froning}, {Linsky}, {Roberge},
  {Stocke}, {Tian}, {Bushinsky}, {D{\'e}sert}, {Mauas}, {Vieytes}, \&
  {Walkowicz}}]{france13}
{France}, K., {Froning}, C.~S., {Linsky}, J.~L., {et~al.} 2013, \apj, 763, 149,
  \dodoi{10.1088/0004-637X/763/2/149}

\bibitem[{{Gaia Collaboration} {et~al.}(2018){Gaia Collaboration}, {Brown},
  {Vallenari}, {Prusti}, {de Bruijne}, {Babusiaux}, {Bailer-Jones}, {Biermann},
  {Evans}, {Eyer}, \& et~al.}]{gaiadr2}
{Gaia Collaboration}, {Brown}, A.~G.~A., {Vallenari}, A., {et~al.} 2018, \aap,
  616, A1, \dodoi{10.1051/0004-6361/201833051}

\bibitem[{{Gershberg}(1972)}]{gershberg72}
{Gershberg}, R.~E. 1972, \apss, 19, 75, \dodoi{10.1007/BF00643168}

\bibitem[{{Gershberg}(2005)}]{gershberg05}
---. 2005, {Solar-Type Activity in Main-Sequence Stars},
  \dodoi{10.1007/3-540-28243-2}

\bibitem[{{Gizis} {et~al.}(2000){Gizis}, {Monet}, {Reid}, {Kirkpatrick},
  {Liebert}, \& {Williams}}]{gizis00}
{Gizis}, J.~E., {Monet}, D.~G., {Reid}, I.~N., {et~al.} 2000, \aj, 120, 1085,
  \dodoi{10.1086/301456}

\bibitem[{{Gizis} {et~al.}(2017){Gizis}, {Paudel}, {Schmidt}, {Williams}, \&
  {Burgasser}}]{gizis17}
{Gizis}, J.~E., {Paudel}, R.~R., {Schmidt}, S.~J., {Williams}, P.~K.~G., \&
  {Burgasser}, A.~J. 2017, \apj, 838, 22, \dodoi{10.3847/1538-4357/aa6197}

\bibitem[{{G{\"u}nther} {et~al.}(2019){G{\"u}nther}, {Zhan}, {Seager},
  {Rimmer}, {Ranjan}, {Stassun}, {Oelkers}, {Daylan}, {Newton}, {Gillen},
  {Rappaport}, {Ricker}, {Latham}, {Winn}, {Jenkins}, {Glidden}, {Fausnaugh},
  {Levine}, {Dittmann}, {Quinn}, {Krishnamurthy}, \& {Ting}}]{gunther19}
{G{\"u}nther}, M.~N., {Zhan}, Z., {Seager}, S., {et~al.} 2019, arXiv e-prints,
  arXiv:1901.00443.
\newblock \doarXiv{1901.00443}

\bibitem[{{Hawley} {et~al.}(2014){Hawley}, {Davenport}, {Kowalski},
  {Wisniewski}, {Hebb}, {Deitrick}, \& {Hilton}}]{hawley14}
{Hawley}, S.~L., {Davenport}, J.~R.~A., {Kowalski}, A.~F., {et~al.} 2014, \apj,
  797, 121, \dodoi{10.1088/0004-637X/797/2/121}

\bibitem[{{Hawley} {et~al.}(1996){Hawley}, {Gizis}, \& {Reid}}]{hawley96}
{Hawley}, S.~L., {Gizis}, J.~E., \& {Reid}, I.~N. 1996, \aj, 112, 2799,
  \dodoi{10.1086/118222}

\bibitem[{{Hilton} {et~al.}(2011){Hilton}, {Hawley}, {Kowalski}, \&
  {Holtzman}}]{hilton11}
{Hilton}, E.~J., {Hawley}, S.~L., {Kowalski}, A.~F., \& {Holtzman}, J. 2011, in
  Astronomical Society of the Pacific Conference Series, Vol. 448, 16th
  Cambridge Workshop on Cool Stars, Stellar Systems, and the Sun, ed.
  C.~{Johns-Krull}, M.~K. {Browning}, \& A.~A. {West}, 197

\bibitem[{{Howard} {et~al.}(2019){Howard}, {Corbett}, {Law}, {Ratzloff},
  {Glazier}, {Fors}, {del Ser}, \& {Haislip}}]{howard:2019}
{Howard}, W.~S., {Corbett}, H., {Law}, N.~M., {et~al.} 2019, \apj, 881, 9,
  \dodoi{10.3847/1538-4357/ab2767}

\bibitem[{{Howard} {et~al.}(2018){Howard}, {Tilley}, {Corbett}, {Youngblood},
  {Loyd}, {Ratzloff}, {Law}, {Fors}, {del Ser}, {Shkolnik}, {Ziegler}, {Goeke},
  {Pietraallo}, \& {Haislip}}]{howard18}
{Howard}, W.~S., {Tilley}, M.~A., {Corbett}, H., {et~al.} 2018, \apjl, 860,
  L30, \dodoi{10.3847/2041-8213/aacaf3}

\bibitem[{{Ilin} {et~al.}(2018){Ilin}, {Schmidt}, {Davenport}, \&
  {Strassmeier}}]{ilin18}
{Ilin}, E., {Schmidt}, S.~J., {Davenport}, J.~R.~A., \& {Strassmeier}, K.~G.
  2018, in 20th Cambridge Workshop on Cool Stars, Stellar Systems and the Sun,
  1

\bibitem[{{Jayasinghe} {et~al.}(2018){Jayasinghe}, {Kochanek}, {Stanek},
  {Shappee}, {Holoien}, {Thompson}, {Prieto}, {Dong}, {Pawlak}, {Shields},
  {Pojmanski}, {Otero}, {Britt}, \& {Will}}]{jayasinghe18}
{Jayasinghe}, T., {Kochanek}, C.~S., {Stanek}, K.~Z., {et~al.} 2018, \mnras,
  477, 3145, \dodoi{10.1093/mnras/sty838}

\bibitem[{{Jayasinghe} {et~al.}(2019){Jayasinghe}, {Stanek}, {Kochanek},
  {Shappee}, {Holoien}, {Thompson}, {Prieto}, {Dong}, {Pawlak}, {Pejcha},
  {Shields}, {Pojmanski}, {Otero}, {Britt}, \& {Will}}]{2019MNRAS.486.1907J}
{Jayasinghe}, T., {Stanek}, K.~Z., {Kochanek}, C.~S., {et~al.} 2019, \mnras,
  486, 1907, \dodoi{10.1093/mnras/stz844}

\bibitem[{{Jenkins} {et~al.}(2002){Jenkins}, {Caldwell}, \&
  {Borucki}}]{jenkins2002}
{Jenkins}, J.~M., {Caldwell}, D.~A., \& {Borucki}, W.~J. 2002, \apj, 564, 495,
  \dodoi{10.1086/324143}

\bibitem[{{Jones} \& {West}(2016)}]{jones:2016}
{Jones}, D.~O., \& {West}, A.~A. 2016, \apj, 817, 1,
  \dodoi{10.3847/0004-637X/817/1/1}

\bibitem[{{Kaiser}(2004)}]{kaiser04}
{Kaiser}, N. 2004, in \procspie, Vol. 5489, Ground-based Telescopes, ed. J.~M.
  {Oschmann}, Jr., 11--22

\bibitem[{{Kiraga} \& {Stepien}(2007)}]{kiraga07}
{Kiraga}, M., \& {Stepien}, K. 2007, \actaa, 57, 149.
\newblock \doarXiv{0707.2577}

\bibitem[{{Kochanek} {et~al.}(2017){Kochanek}, {Shappee}, {Stanek}, {Holoien},
  {Thompson}, {Prieto}, {Dong}, {Shields}, {Will}, {Britt}, {Perzanowski}, \&
  {Pojma{\'n}ski}}]{kochanek17}
{Kochanek}, C.~S., {Shappee}, B.~J., {Stanek}, K.~Z., {et~al.} 2017, \pasp,
  129, 104502, \dodoi{10.1088/1538-3873/aa80d9}

\bibitem[{{Kov{\'a}cs} {et~al.}(2002){Kov{\'a}cs}, {Zucker}, \&
  {Mazeh}}]{2002A&A...391..369K}
{Kov{\'a}cs}, G., {Zucker}, S., \& {Mazeh}, T. 2002, \aap, 391, 369,
  \dodoi{10.1051/0004-6361:20020802}

\bibitem[{{Kowalski} {et~al.}(2009){Kowalski}, {Hawley}, {Hilton}, {Becker},
  {West}, {Bochanski}, \& {Sesar}}]{kowalski09}
{Kowalski}, A.~F., {Hawley}, S.~L., {Hilton}, E.~J., {et~al.} 2009, \aj, 138,
  633, \dodoi{10.1088/0004-6256/138/2/633}

\bibitem[{{Kowalski} {et~al.}(2010){Kowalski}, {Hawley}, {Holtzman},
  {Wisniewski}, \& {Hilton}}]{kowalski10}
{Kowalski}, A.~F., {Hawley}, S.~L., {Holtzman}, J.~A., {Wisniewski}, J.~P., \&
  {Hilton}, E.~J. 2010, \apjl, 714, L98, \dodoi{10.1088/2041-8205/714/1/L98}

\bibitem[{{Kowalski} {et~al.}(2013){Kowalski}, {Hawley}, {Wisniewski}, {Osten},
  {Hilton}, {Holtzman}, {Schmidt}, \& {Davenport}}]{kowalski13}
{Kowalski}, A.~F., {Hawley}, S.~L., {Wisniewski}, J.~P., {et~al.} 2013, \apjs,
  207, 15, \dodoi{10.1088/0067-0049/207/1/15}

\bibitem[{{Lacy} {et~al.}(1976){Lacy}, {Moffett}, \& {Evans}}]{lacy76}
{Lacy}, C.~H., {Moffett}, T.~J., \& {Evans}, D.~S. 1976, \apjs, 30, 85,
  \dodoi{10.1086/190358}

\bibitem[{{Lafler} \& {Kinman}(1965)}]{1965ApJS...11..216L}
{Lafler}, J., \& {Kinman}, T.~D. 1965, \apjs, 11, 216, \dodoi{10.1086/190116}

\bibitem[{{Lammer} {et~al.}(2007){Lammer}, {Lichtenegger}, {Kulikov},
  {Grie{\ss}meier}, {Terada}, {Erkaev}, {Biernat}, {Khodachenko}, {Ribas},
  {Penz}, \& {Selsis}}]{lammer07}
{Lammer}, H., {Lichtenegger}, H.~I.~M., {Kulikov}, Y.~N., {et~al.} 2007,
  Astrobiology, 7, 185, \dodoi{10.1089/ast.2006.0128}

\bibitem[{{Law} {et~al.}(2009){Law}, {Kulkarni}, {Dekany}, {Ofek}, {Quimby},
  {Nugent}, {Surace}, {Grillmair}, {Bloom}, {Kasliwal}, {Bildsten}, {Brown},
  {Cenko}, {Ciardi}, {Croner}, {Djorgovski}, {van Eyken}, {Filippenko}, {Fox},
  {Gal-Yam}, {Hale}, {Hamam}, {Helou}, {Henning}, {Howell}, {Jacobsen},
  {Laher}, {Mattingly}, {McKenna}, {Pickles}, {Poznanski}, {Rahmer}, {Rau},
  {Rosing}, {Shara}, {Smith}, {Starr}, {Sullivan}, {Velur}, {Walters}, \&
  {Zolkower}}]{law09}
{Law}, N.~M., {Kulkarni}, S.~R., {Dekany}, R.~G., {et~al.} 2009, \pasp, 121,
  1395, \dodoi{10.1086/648598}

\bibitem[{{L{\'e}pine} {et~al.}(2013){L{\'e}pine}, {Hilton}, {Mann}, {Wilde},
  {Rojas-Ayala}, {Cruz}, \& {Gaidos}}]{lepine13}
{L{\'e}pine}, S., {Hilton}, E.~J., {Mann}, A.~W., {et~al.} 2013, \aj, 145, 102,
  \dodoi{10.1088/0004-6256/145/4/102}

\bibitem[{{Loyd} {et~al.}(2018){Loyd}, {France}, {Youngblood}, {Schneider},
  {Brown}, {Hu}, {Segura}, {Linsky}, {Redfield}, {Tian}, {Rugheimer}, {Miguel},
  \& {Froning}}]{loyd18}
{Loyd}, R.~O.~P., {France}, K., {Youngblood}, A., {et~al.} 2018, \apj, 867, 71,
  \dodoi{10.3847/1538-4357/aae2bd}

\bibitem[{{Luger} \& {Barnes}(2015)}]{luger15}
{Luger}, R., \& {Barnes}, R. 2015, Astrobiology, 15, 119,
  \dodoi{10.1089/ast.2014.1231}

\bibitem[{{Mondrik} {et~al.}(2018){Mondrik}, {Newton}, \& {Irwin}}]{mondrik18}
{Mondrik}, N., {Newton}, E., \& {Irwin}, D.~C.~J. 2018, ArXiv e-prints.
\newblock \doarXiv{1809.09177}

\bibitem[{{Morales} {et~al.}(2009){Morales}, {Ribas}, {Jordi}, {Torres},
  {Gallardo}, {Guinan}, {Charbonneau}, {Wolf}, {Latham}, {Anglada-Escud{\'e}},
  {Bradstreet}, {Everett}, {O'Donovan}, {Mandushev}, \& {Mathieu}}]{morales09}
{Morales}, J.~C., {Ribas}, I., {Jordi}, C., {et~al.} 2009, \apj, 691, 1400,
  \dodoi{10.1088/0004-637X/691/2/1400}

\bibitem[{{Newton} {et~al.}(2017){Newton}, {Irwin}, {Charbonneau}, {Berlind},
  {Calkins}, \& {Mink}}]{newton17}
{Newton}, E.~R., {Irwin}, J., {Charbonneau}, D., {et~al.} 2017, \apj, 834, 85,
  \dodoi{10.3847/1538-4357/834/1/85}

\bibitem[{{Newton} {et~al.}(2016){Newton}, {Irwin}, {Charbonneau},
  {Berta-Thompson}, {Dittmann}, \& {West}}]{newton16}
---. 2016, \apj, 821, 93, \dodoi{10.3847/0004-637X/821/2/93}

\bibitem[{{O'Malley-James} \& {Kaltenegger}(2019)}]{omalley18}
{O'Malley-James}, J.~T., \& {Kaltenegger}, L. 2019, \mnras, 485, 5598,
  \dodoi{10.1093/mnras/stz724}

\bibitem[{{Paudel} {et~al.}(2018){Paudel}, {Gizis}, {Mullan}, {Schmidt},
  {Burgasser}, {Williams}, \& {Berger}}]{paudel18}
{Paudel}, R.~R., {Gizis}, J.~E., {Mullan}, D.~J., {et~al.} 2018, \apj, 858, 55,
  \dodoi{10.3847/1538-4357/aab8fe}

\bibitem[{{Paudel} {et~al.}(2019){Paudel}, {Gizis}, {Mullan}, {Schmidt},
  {Burgasser}, {Williams}, {Youngblood}, \& {Stassun}}]{paudel19}
---. 2019, \mnras, 486, 1438, \dodoi{10.1093/mnras/stz886}

\bibitem[{{Pojmanski}(1997)}]{pojmanski97}
{Pojmanski}, G. 1997, \actaa, 47, 467

\bibitem[{{Poppenhaeger}(2015)}]{poppenhaeger15}
{Poppenhaeger}, K. 2015, in European Physical Journal Web of Conferences, Vol.
  101, European Physical Journal Web of Conferences, 05002

\bibitem[{{Ricker} {et~al.}(2015){Ricker}, {Winn}, {Vanderspek}, {Latham},
  {Bakos}, {Bean}, {Berta-Thompson}, {Brown}, {Buchhave}, {Butler}, {Butler},
  {Chaplin}, {Charbonneau}, {Christensen-Dalsgaard}, {Clampin}, {Deming},
  {Doty}, {De Lee}, {Dressing}, {Dunham}, {Endl}, {Fressin}, {Ge}, {Henning},
  {Holman}, {Howard}, {Ida}, {Jenkins}, {Jernigan}, {Johnson}, {Kaltenegger},
  {Kawai}, {Kjeldsen}, {Laughlin}, {Levine}, {Lin}, {Lissauer}, {MacQueen},
  {Marcy}, {McCullough}, {Morton}, {Narita}, {Paegert}, {Palle}, {Pepe},
  {Pepper}, {Quirrenbach}, {Rinehart}, {Sasselov}, {Sato}, {Seager},
  {Sozzetti}, {Stassun}, {Sullivan}, {Szentgyorgyi}, {Torres}, {Udry}, \&
  {Villasenor}}]{ricker15}
{Ricker}, G.~R., {Winn}, J.~N., {Vanderspek}, R., {et~al.} 2015, Journal of
  Astronomical Telescopes, Instruments, and Systems, 1, 014003,
  \dodoi{10.1117/1.JATIS.1.1.014003}

\bibitem[{{Rodr{\'{\i}}guez} {et~al.}(2018){Rodr{\'{\i}}guez}, {Schmidt},
  {Jayasinghe}, {Stanek}, {Prieto}, {Shappee}, {Kochanek}, {Thompson},
  {Shields}, {Holoien}, {Bersier}, \& {Brimacombe}}]{rodriguez18}
{Rodr{\'{\i}}guez}, R., {Schmidt}, S.~J., {Jayasinghe}, T., {et~al.} 2018,
  Research Notes of the American Astronomical Society, 2, 8,
  \dodoi{10.3847/2515-5172/aabe7d}

\bibitem[{{Samus'} {et~al.}(2003){Samus'}, {Goranskii}, {Durlevich}, {Zharova},
  {Kazarovets}, {Kireeva}, {Pastukhova}, {Williams}, \& {Hazen}}]{samus:2003}
{Samus'}, N.~N., {Goranskii}, V.~P., {Durlevich}, O.~V., {et~al.} 2003,
  Astronomy Letters, 29, 468, \dodoi{10.1134/1.1589864}

\bibitem[{{Scargle}(1982)}]{1982ApJ...263..835S}
{Scargle}, J.~D. 1982, \apj, 263, 835, \dodoi{10.1086/160554}

\bibitem[{{Schmidt} {et~al.}(2014){Schmidt}, {Prieto}, {Stanek}, {Shappee},
  {Morrell}, {Bardalez Gagliuffi}, {Kochanek}, {Jencson}, {Holoien}, {Basu},
  {Beacom}, {Szczygie{\l}}, {Pojmanski}, {Brimacombe}, {Dubberley}, {Elphick},
  {Foale}, {Hawkins}, {Mullins}, {Rosing}, {Ross}, \& {Walker}}]{schmidt14}
{Schmidt}, S.~J., {Prieto}, J.~L., {Stanek}, K.~Z., {et~al.} 2014, \apjl, 781,
  L24, \dodoi{10.1088/2041-8205/781/2/L24}

\bibitem[{{Schmidt} {et~al.}(2016){Schmidt}, {Shappee}, {Gagn{\'e}}, {Stanek},
  {Prieto}, {Holoien}, {Kochanek}, {Chomiuk}, {Dong}, {Seibert}, \&
  {Strader}}]{schmidt16}
{Schmidt}, S.~J., {Shappee}, B.~J., {Gagn{\'e}}, J., {et~al.} 2016, \apjl, 828,
  L22, \dodoi{10.3847/2041-8205/828/2/L22}

\bibitem[{{Schmidt} {et~al.}(2018){Schmidt}, {Shappee}, {van Saders}, {Stanek},
  {Brown}, {Kochanek}, {Dong}, {Drout}, {Frank}, {Holoien}, {Johnson},
  {Madore}, {Prieto}, {Seibert}, {Seidel}, \& {Simonian}}]{schmidt18}
{Schmidt}, S.~J., {Shappee}, B.~J., {van Saders}, J.~L., {et~al.} 2018, ArXiv
  e-prints.
\newblock \doarXiv{1809.04510}

\bibitem[{{Segura} {et~al.}(2010){Segura}, {Walkowicz}, {Meadows}, {Kasting},
  \& {Hawley}}]{segura10}
{Segura}, A., {Walkowicz}, L.~M., {Meadows}, V., {Kasting}, J., \& {Hawley}, S.
  2010, Astrobiology, 10, 751, \dodoi{10.1089/ast.2009.0376}

\bibitem[{{Shappee} {et~al.}(2014){Shappee}, {Prieto}, {Grupe}, {Kochanek},
  {Stanek}, {De Rosa}, {Mathur}, {Zu}, {Peterson}, {Pogge}, {Komossa}, {Im},
  {Jencson}, {Holoien}, {Basu}, {Beacom}, {Szczygie{\l}}, {Brimacombe},
  {Adams}, {Campillay}, {Choi}, {Contreras}, {Dietrich}, {Dubberley},
  {Elphick}, {Foale}, {Giustini}, {Gonzalez}, {Hawkins}, {Howell}, {Hsiao},
  {Koss}, {Leighly}, {Morrell}, {Mudd}, {Mullins}, {Nugent}, {Parrent},
  {Phillips}, {Pojmanski}, {Rosing}, {Ross}, {Sand}, {Terndrup}, {Valenti},
  {Walker}, \& {Yoon}}]{shappee14}
{Shappee}, B.~J., {Prieto}, J.~L., {Grupe}, D., {et~al.} 2014, \apj, 788, 48,
  \dodoi{10.1088/0004-637X/788/1/48}

\bibitem[{{Shields} {et~al.}(2016){Shields}, {Ballard}, \&
  {Johnson}}]{shields16}
{Shields}, A.~L., {Ballard}, S., \& {Johnson}, J.~A. 2016, \physrep, 663, 1,
  \dodoi{10.1016/j.physrep.2016.10.003}

\bibitem[{{Simonian} {et~al.}(2016){Simonian}, {Stanek}, {Schmidt}, {Kochanek},
  {Brown}, {Holoien}, {Godoy-Rivera}, {Basu}, {Shappee}, {Prieto}, {Bersier},
  {Dong}, {Chen}, \& {Brimacombe}}]{simonian16}
{Simonian}, G., {Stanek}, K.~Z., {Schmidt}, S., {et~al.} 2016, The Astronomer's
  Telegram, 8803

\bibitem[{{Stanek} {et~al.}(2013){Stanek}, {Shappee}, {Kochanek}, {Holoien},
  {Jencson}, {Basu}, {Beacom}, {Prieto}, {Szczygiel}, {Pojmanski}, {Dubberley},
  {Elphick}, {Foale}, {Hawkins}, {Mullens}, {Rosing}, {Ross}, {Walker}, \&
  {Brimacombe}}]{stanek13}
{Stanek}, K.~Z., {Shappee}, B.~J., {Kochanek}, C.~S., {et~al.} 2013, The
  Astronomer's Telegram, 5276

\bibitem[{{Tilley} {et~al.}(2017){Tilley}, {Segura}, {Meadows}, {Hawley}, \&
  {Davenport}}]{tilley17}
{Tilley}, M.~A., {Segura}, A., {Meadows}, V.~S., {Hawley}, S., \& {Davenport},
  J. 2017, arXiv e-prints, arXiv:1711.08484.
\newblock \doarXiv{1711.08484}

\bibitem[{{Tyson}(2002)}]{tyson2002}
{Tyson}, J.~A. 2002, in \procspie, Vol. 4836, Survey and Other Telescope
  Technologies and Discoveries, ed. J.~A. {Tyson} \& S.~{Wolff}, 10--20

\bibitem[{{Vida} {et~al.}(2017){Vida}, {K{\H o}v{\'a}ri}, {P{\'a}l},
  {Ol{\'a}h}, \& {Kriskovics}}]{vida17}
{Vida}, K., {K{\H o}v{\'a}ri}, Z., {P{\'a}l}, A., {Ol{\'a}h}, K., \&
  {Kriskovics}, L. 2017, \apj, 841, 124, \dodoi{10.3847/1538-4357/aa6f05}

\bibitem[{{Vida} {et~al.}(2013){Vida}, {Kriskovics}, \& {Ol{\'a}h}}]{vida13}
{Vida}, K., {Kriskovics}, L., \& {Ol{\'a}h}, K. 2013, Astronomische
  Nachrichten, 334, 972, \dodoi{10.1002/asna.201211973}

\bibitem[{{Vida} {et~al.}(2014){Vida}, {Ol{\'a}h}, \& {Szab{\'o}}}]{vida14}
{Vida}, K., {Ol{\'a}h}, K., \& {Szab{\'o}}, R. 2014, \mnras, 441, 2744,
  \dodoi{10.1093/mnras/stu760}

\bibitem[{{Walkowicz} \& {Hawley}(2009)}]{walkowicz09}
{Walkowicz}, L.~M., \& {Hawley}, S.~L. 2009, \aj, 137, 3297,
  \dodoi{10.1088/0004-6256/137/2/3297}

\bibitem[{{Watson} {et~al.}(2006){Watson}, {Henden}, \& {Price}}]{watson06}
{Watson}, C.~L., {Henden}, A.~A., \& {Price}, A. 2006, Society for Astronomical
  Sciences Annual Symposium, 25, 47

\bibitem[{{Wenger} {et~al.}(2000){Wenger}, {Ochsenbein}, {Egret}, {Dubois},
  {Bonnarel}, {Borde}, {Genova}, {Jasniewicz}, {Lalo{\"e}}, {Lesteven}, \&
  {Monier}}]{wenger:2000}
{Wenger}, M., {Ochsenbein}, F., {Egret}, D., {et~al.} 2000, \aaps, 143, 9,
  \dodoi{10.1051/aas:2000332}

\bibitem[{{West} {et~al.}(2006){West}, {Bochanski}, {Hawley}, {Cruz}, {Covey},
  {Silvestri}, {Reid}, \& {Liebert}}]{west2006}
{West}, A.~A., {Bochanski}, J.~J., {Hawley}, S.~L., {et~al.} 2006, \aj, 132,
  2507, \dodoi{10.1086/508652}

\bibitem[{{West} \& {Hawley}(2008)}]{west08}
{West}, A.~A., \& {Hawley}, S.~L. 2008, \pasp, 120, 1161,
  \dodoi{10.1086/593024}

\bibitem[{{West} {et~al.}(2008){West}, {Hawley}, {Bochanski}, {Covey}, {Reid},
  {Dhital}, {Hilton}, \& {Masuda}}]{west2008}
{West}, A.~A., {Hawley}, S.~L., {Bochanski}, J.~J., {et~al.} 2008, \aj, 135,
  785, \dodoi{10.1088/0004-6256/135/3/785}

\bibitem[{{West} {et~al.}(2004){West}, {Hawley}, {Walkowicz}, {Covey},
  {Silvestri}, {Raymond}, {Harris}, {Munn}, {McGehee}, {Ivezi{\'c}}, \&
  {Brinkmann}}]{west:2004}
{West}, A.~A., {Hawley}, S.~L., {Walkowicz}, L.~M., {et~al.} 2004, \aj, 128,
  426, \dodoi{10.1086/421364}

\bibitem[{{West} {et~al.}(2011){West}, {Morgan}, {Bochanski}, {Andersen},
  {Bell}, {Kowalski}, {Davenport}, {Hawley}, {Schmidt}, {Bernat}, {Hilton},
  {Muirhead}, {Covey}, {Rojas-Ayala}, {Schlawin}, {Gooding}, {Schluns},
  {Dhital}, {Pineda}, \& {Jones}}]{west11}
{West}, A.~A., {Morgan}, D.~P., {Bochanski}, J.~J., {et~al.} 2011, \aj, 141,
  97, \dodoi{10.1088/0004-6256/141/3/97}

\bibitem[{{Yang} {et~al.}(2017){Yang}, {Liu}, {Gao}, {Fang}, {Guo}, {Zhang},
  {Hou}, {Wang}, \& {Cao}}]{yang:2017}
{Yang}, H., {Liu}, J., {Gao}, Q., {et~al.} 2017, \apj, 849, 36,
  \dodoi{10.3847/1538-4357/aa8ea2}

\bibitem[{{Zechmeister} \& {K{\"u}rster}(2009)}]{2009A&A...496..577Z}
{Zechmeister}, M., \& {K{\"u}rster}, M. 2009, \aap, 496, 577,
  \dodoi{10.1051/0004-6361:200811296}

\end{thebibliography}

\end{document}